\documentclass[12pt, oneside]{article}
\usepackage{epsfig}
\usepackage{latexsym}
\usepackage{amssymb}
\usepackage{amsmath}
\usepackage{graphicx}
\usepackage{subfigure} 
\usepackage{amsfonts}
\usepackage{array}
\usepackage{color}
\usepackage{fancyhdr}
\fancypagestyle{plain}{% this is for beginning of chapter: empty
 \fancyhead{} % get rid of headers 
 }

\makeatletter
\def\cleardoublepage{\clearpage\if@twoside \ifodd\c@page\else%
	     \hbox{}%
	 \thispagestyle{empty}%              % Empty header styles
	 \newpage%
	 \if@twocolumn\hbox{}\newpage\fi\fi\fi}
\makeatother
\numberwithin{equation}{section}

\newcommand{\be}{\begin{equation}}
\newcommand{\ee}{\end{equation}}
\newcommand{\bea}{\begin{eqnarray}}
\newcommand{\eea}{\end{eqnarray}}

\newcommand{\mn}{{\mu\nu}}

\newcommand{\psip}{{\psi^\prime}}

\newcommand{\Pis}{\Pi_\psi}
\newcommand{\sub}{\left( \frac{N_r}{N} \right)}

\newcommand{\m}{\mathcal{M}}
\newcommand{\R}{\mathcal{R}}

\def\gsim{\raise0.3ex\hbox{$>$\kern-0.75em\raise-1.1ex\hbox{$\sim$}}}
\def\lsim{\raise0.3ex\hbox{$<$\kern-0.75em\raise-1.1ex\hbox{$\sim$}}}

\newcommand{\Tm}{{T^\mu}_{\mu \ max}}

\begin{document}

\title{\textbf{Critical Collapse in Einstein-Gauss-Bonnet Gravity in Five and
Six Dimensions}}

\author{{\bf N. Deppe$^a$,  C. D. Leonard$^{c,d}$, T. Taves$^{b, f}$ }\\
{\bf G. Kunstatter$^{a, b}$, R.B. Mann$^{c,e}$}\\[10pt]
{\small $^a$ Department of Physics and Astronomy, University of Winnipeg}\\
{\small Winnipeg, Manitoba, R3B 2E9, Canada}\\[5pt]
{\small $^b$Winnipeg Institute of Theoretical Physics, University of Winnipeg}
\\[5pt] 
{\small $^c$ Department of Physics and Astronomy, University of Waterloo}\\
{\small Waterloo, Ontario, N2L 3G1, Canada}\\[5pt]
{\small $^d$ Department of Physics and Physical Oceanography, Memorial
University of Newfoundland}\\
{\small St. John's, Newfoundland, A1B 3X7, Canada}\\[5pt]
{\small $^e$ Perimeter Institute, 31 Caroline Street North, Waterloo, Ontario,
N2L 2Y5, Canada}\\[5pt]
{\small $^f$ Department of Physics and Astronomy, University of Manitoba}\\
{\small Winnipeg, Manitoba, R3T 2N2 Canada}
 }
 
 \date{\today}

\maketitle

\begin{abstract}

Einstein-Gauss-Bonnet gravity (EGB) provides a natural higher dimensional and higher order curvature generalization of Einstein gravity. It contains a new, presumably microscopic, length scale that should affect short distance properties of the dynamics, such as Choptuik scaling. We present the results of a numerical analysis in generalized flat slice co-ordinates of 
%to the action of general relativity 
self-gravitating massless scalar spherical collapse in five and six dimensional EGB gravity near the threshold of black hole formation.  %Although the non-scale invariant nature of Einstein-Gauss-Bonnet gravity destroys the discrete self similarity, we nonetheless 
Remarkably, the behaviour is universal (i.e. independent of initial data) but qualitatively different in five and six dimensions. In five dimensions there is a minimum horizon radius, suggestive of a first order transition between black hole and dispersive initial data. In six dimensions no radius gap is evident. Instead, below the GB scale there is a change in the critical exponent and echoing period. 
% and the appearence of locally naked singularity.

\end{abstract}

\clearpage

\section{Introduction}
\label{Intro}

Recent interest in string theory has popularized the study of higher dimensional
and higher curvature gravity. The Einstein action has many desirable properties: it is second order in derivatives of the metric, ghost free when linearized about a flat background and obeys a Birkhoff theorem that yields a one parameter family of spherically symmetric black hole solutions. In four dimensions the Ricci scalar is the only curvature invariant with these properties but in dimensions greater than four it is possible to add higher order curvature terms in the form of so-called Lovelock polynomials
 \cite{Lanczos1938, Lovelock1970, Lovelock1971}. These terms contribute to the equations of motion
while retaining no more than second derivatives of the metric. Moreover, they have been proven to be ghost-free \cite{ghost free} and obey a generalized Birkhoff theorem  \cite{gen birkhoff, deser birkhoff}. 

The Lovelock
action, $I$, written in terms of the Lovelock polynomials, ${\cal L}_{(p)}$ is
given by
\be
I=\frac{1}{2\kappa_n^2}\int d^nx\sqrt{-g}\sum^{[n/2]}_{p=0}\alpha_{(p)} {\cal
L}_{(p)},
\label{eq:lovelock action}
\ee
\be
{\cal L}_{(p)}=
\frac{p!}{2^p}\delta^{\mu_1...\mu_p\nu_1..\nu_p}_{
\rho_1...\rho_p\sigma_1..\sigma_p}
    \R_{\mu_1\nu_1}{}^{\rho_1\sigma_1}...\R_{\mu_p\nu_p}{}^{\rho_p\sigma_p},
 \label{eq:lovelock lagrangian}
\ee
where $\kappa_n = \sqrt{8 \pi G_n}$, $G_n$ is Newton's gravitational constant, $g$ is the determinant of the metric, $n$ is the number of spacetime dimensions, $[n/2]$ refers to the largest integer less than or equal to $n/2$,  $\delta^{\mu_1...\mu_p}_{\rho_1...\rho_p}:=\delta^{\mu_1}_{[p_1}...\delta^{\mu_p }_{p_p]}\,$ and $\R_{\mu \nu \rho \sigma}$ is the Riemann curvature tensor. 
$\alpha_{(p)}$ are coupling constants of dimension (length) ${}^{2(p-1)}$. 
The first two terms, ${\cal L}_{(0)}$ and ${\cal L}_{(1)}$, correspond to the
cosmological constant and Einstein-Hilbert term, respectively, while  ${\cal L}_{(2)}$ is the Gauss-Bonnet (GB) term.  It has been argued \cite{string lovelock} that the 
GB term appears in the low-energy limit for strings propagating in curved spacetime. Here we focus on the simplest non-trivial theory, namely Einstein-Gauss-Bonnet (EGB) gravity, containing only the Einstein term and $p=2$ GB term \footnote{The addition of a cosmological constant should not affect the short distance behaviour that is the subject of our paper.}.

%General relativity has made predictions which agree with experiments on
%cosmological and astronomical scales, therefore the Lovelock polynomials can
%only contribute to predictions which would be made on smaller than astronomical
%scales such as the collapse of microscopic black holes.  

It has been known for quite some time that the spherically symmetric collapse of a
massless scalar field minimally coupled to GR exhibits critical behaviour   \cite{Choptuik1993}.
Specifically, numerical studies of black hole formation
indicate that for any parameter in the initial data, $A$ say, there exists a corresponding critical
value $A^*$ such that for $A>A^*$ a black hole forms while for $A<A^*$ the matter 
disperses to infinity.  Black holes with $A$ just slightly bigger than $A^*$ are
known as near critical black holes.  Near criticality all geometrical quantities describing the black hole, such as its mass, obey a scaling relation of the form:
\be
\ln(M_{BH}) = \gamma  \ln(A-A^*) + f(A-A^*)
\label{CS}
\ee
where $f$ is a periodic function of its argument. The critical exponent, $\gamma$ and the period $T$ of 
$f$, are universal\footnote{$\gamma$ depends only on the scaling dimension of the quantity considered. For example in 4D $M$ scales as length so the radius and mass of the black hole have the same critical exponent. Curvature, on the other hand scales as (length)$^2$, so its critical exponent is double that of mass.} in the sense that they are independent of the form of the initial data or the specifics of the parameter $A$ that is varied. This universality and the vanishing of $M_{BH}$ at criticality (cf. Eq.(\ref{CS}))
 suggest a second order phase transition between the black hole and dispersive end states of the collapse. This fascinating behaviour was ultimately explained using renormalization group arguments in the context of radiation fluids\cite{Koike}. For spherical massless scalar field collapse, the critical exponent  and echoing period were obtained by Gundlach \cite{Gundlach1997} from the properties of a critical, discretely self-similar zero mass black hole solution that behaves like an intermediate attractor in the space of solutions. 

The critical exponent $\gamma$ and period $T$ do  depend on the number of dimensions and the type of matter. In addition, the form of $f$ can depend on the
space-time slicing and the particular quantity that is being measured. For example, in Schwarzschild and null co-ordinates the radius of the horizon on formation yields an $f$ that 
is well fit to a small amplitude sine wave. 
By comparison, in flat slice or Painlev\'{e}-Gullstrand (PG)
co-ordinates the periodic function that describes the scaling of the apparent horizon on formation exhibits large amplitude cusps \cite{Ziprick2009, Ziprick2009b}. The difference can be understood by noting that in the former case one is measuring a quantity very close to the radius of the final event horizon, whereas in PG co-ordinates the apparent horizon is detected much earlier. The large amplitude cusps are likely due to that fact that near criticality in PG co-ordinates the apparent horizon forms at small radius and hence in the strong field region where such large fluctuations are expected\footnote{We are grateful to Patrick Brady for suggesting this  explanation.}. On the other hand the scaling relation for the
maximum value of the Ricci scalar at the origin for subcritical
evolution is invariant and exhibits slicing independent small oscillations. %Collectively mass scaling and Ricci scaling both fall into the category of Choptuik scaling which is attributed to the discrete self similarity (DSS) of the solution near criticality.  

The presence of a dimensionful constant in the action in general changes the above scenario, as verified for Yang-Mills collapse  \cite{Choptuik1996}, massive scalar field collapse \cite{Brady1997} and massive gauge field collapse \cite{Garfinkle:2003jf}.
 In massive scalar field collapse, for initial data whose width is smaller than the Compton wavelength of the scalar field, the usual second order phase transition is found, whereas in the other limit the phase transition exhibits a mass gap and is first order. It is clearly of interest to study the effects on Choptuik scaling of the Gauss-Bonnet parameter and higher order Lovelock coupling constants. Golod and Piran \cite{Golod2012} recently presented such an analysis for the spherical collapse of massless scalar matter coupled to EGB gravity in five
dimensions using double null co-ordinates. They found, as expected, that the Gauss-Bonnet term dominates the dynamics at short distances and destroys the discrete self-similarity characteristic of Choptuik scaling.  Their work concentrated on the regime where the GB terms strongly dominated the dynamics.% and left open questions about the nature of collapse when the GB parameter was large but still small enough that the GR terms still played an important role.

The purpose of the present work is to investigate further the critical collapse of a spherically symmetric, massless scalar field minimally coupled to five and six dimensional EGB gravity. We work in flat slice, or generalized Painlev\'{e}-Gullstrand (PG),
co-ordinates since they have several advantages over double null co-ordinates in the present context:  They are regular at apparent horizons so that the simulations can run up to (and even past) horizon formation.  Hence one can calculate the time and position of horizon formation without having to stop the code at some arbitrary distance before horizon formation as would be necessary in Schwarzschild and null co-ordinates.  More importantly, the cusp-like nature of the horizon scaling function in PG co-ordinates has the advantage of making the potential appearance of the periodicity in an equation such as (\ref{CS}) more obvious.  It should be pointed out that one disadvantage of using PG co-ordinates is the lack of automatic spacial mesh refinement which occurs near horizon formation in null co-ordinates.  As we will explain in the next section, the nature of the dynamical equations suggest that qualitative differences can occur in different numbers of spacetime dimensions. It is for this reason that we investigate both five and six spacetime dimensions.  %We also bridge the gap between collapse in the GR case and collapse in the extreme GB case and show that there is a transition region where a new type of mass and Ricci scaling occur.

%\footnote{We leave a detailed study of higher dimensions for future study.}. Although the qualitative behaviour turns out to be similar in both cases, the numerics is significantly better in six dimensions and produces the best results.

We confirm some of the results in five dimensions \cite{Golod2012}, extend the analysis to six dimensions and obtain some surprising new results in both five and six dimensions. For all initial data  and choice of parameter $A$ that we examined there exists a critical value $A^{*}$ that separates black hole formation from dispersion.  As expected, when the horizon forms far from the singularity  
the GR term dominates and the standard Choptuik critical scaling relation is found. Things change as one gets close enough to criticality to enter the region in which the GB terms dominate the dynamics. Near criticality the scalar field  at the origin oscillates with a constant period $T$ that converges as $(A-A^{*})\to0$ to a value  that depends on the GB parameter as previously shown  \cite{Golod2012}.   We find a different relationship between $T$ and the GB parameter than in \cite{Golod2012}, albeit for smaller values of the GB parameter.
%{\bf Still need to see how this pans out then finish this part - Danielle}  In five dimensions we confirm that $\lambda_c\propto [\alpha_{(2)}]^\beta$ where $\beta_{(5)}\sim0.5$ as expected from dimensional arguments \cite{Golod2012}.  In six dimensions the numerics appear to be more problematic, but to the accuracy we are able to obtain, we surprisingly find a different value of $\beta_{(6)}{\bf \sim???}$.

In addition, we explore in detail the scaling in the GB dominated region. 
We find qualitatively different behaviour in five and six dimensions. In five dimensions there is evidence for a radius gap: in the supercritical region the radius of the apparent horizon on formation asymptotes to a constant value as criticality is approached from above. The maximum value of the trace of the energy momentum tensor\footnote{In a previous version we referred to $\Tm$ as the "maximum value of the Ricci scalar at the origin".  In the GR case this is true but not in the EGB case.  See (2.13) and (2.14) of \cite{Maeda2011} for the relationship between the energy momentum tensor and the curvature invariants in the EGB case.} at the origin, $\Tm$, also appears to approach a constant value as criticality is approached from below. 
%In the supercritical region, in both five and six dimensions, as the black holes get smaller the slope of the apparent horizon radius scaling plot decreases, as does the oscillation period.  In addition, the horizon radius appears to be bounded below, which suggests a radius/mass gap normally associated with a first order phase transition. Remarkably, the shape of the horizon radius scaling plot is univeral: it is independent of the type of initial data and the parameter being varied. This universality also appears in the sub-critical region. 
%In the GB region, the plot of the maximum Ricci scalar  at the origin as a function of $A-A*$ is independent of which parameter $A$ is being varied and also of the form of the initial data. 

In six dimensions, the behaviour is qualitatively different. In the GB region the radius of the apparent horizon formation obeys a relationship similar to (\ref{CS}) but with different exponent and period. $\Tm$ also exhibits this same scaling relation with another scaling exponent, and small, but irregular oscillations.
%Our results suggest the possible existence of an unstable critical solution that is an intermediate attractor in the solution space. If the analogy with the massive scalar field is valid, this solution has finite mass and is non-singular \cite{Brady1997}. 

The rest of this work is organized as follows.  In section \ref{EoM} we describe
the equations of motion which we derived using Hamiltonian formalism
 \cite{Taves2012}.  In section \ref{Numerics and Methods} we discuss the numerical
implementation of the solution and describe the general methods used to obtain results.  In section \ref{Results} we give our results and conclude in section \ref{Conclusion}.

\section{Equations of Motion}
\label{EoM}

As stated above, we start with the action for a massless scalar field $\psi$ minimally coupled to the EGB action:
 \bea
  I= \frac{1}{2 \kappa_n^2}\int d^nx \sqrt{-g}\left(\R + \alpha_{(2)}\left[ \R^2 - 4\R_\mn \R^\mn + \R_{\mu \nu \rho \sigma}\R^{\mu \nu \rho \sigma}   \right]   + \kappa_n^2 \left(\nabla\psi\right)^2
  \right),
  \label{eq:GB action}
  \eea
We use the ADM metric parametrization:
  \be
ds^2= -N^2(x,t)dt^2+\Lambda^2(x,t)(dx+N_r(x,t)dt)^2+R^2(x,t)d\Omega^{2}
\ee
where $R$ is the areal radius. It is also useful to define the Misner-Sharp mass function\cite{Misner_Sharp}, suitably generalized to EGB\cite{Maeda_MS}:
\bea
{\cal M}&:=& \frac{1}{2kG}\left[
R^{n-3}\left(1-(DR)^2\right) + 
\tilde{\alpha}_{(2)}R^{n-5}\left(1-(DR)^2\right)^2 \right]
\label{eq:M EL}
\eea
%In spherical symmetry $|\nabla R|^2$ is proportional to the null expansion and therefore vanishes on apparent horizons. This gives a simple relationship between the radius $R_{ah}$ of the horizon and the mass function $M_{ah}$ at $R_{ah}$.
where the $n$ dimensional gravitational constant, $G$ is defined as $2kG = 2 \kappa_n^2/(n-2)A_{n-2}$  \cite{Ziprick2009}, $A_{n-2}$ is the surface area of an $n-2$ dimensional sphere, $k = 8(n-3)/(n-2)^2$, $D$ is the 2 dimensional covariant derivative and $\tilde{\alpha}_{(2)}:=((n-3)!/(n-5)!)\alpha_{(2)}$.  In the following we work in units in which $2G=1$.  We work in flat slice co-ordinates $x=R$ and $\Lambda=1$ in which the equations of motion for the scalar field and its conjugate momentum are  \cite{Taves2012}:
\be
\label{psidot_new}
\dot{\psi} = N \left( \frac{\Pis}{R^{n-2}} + \left( \frac{N_r}{N} \right)
\psip \right)
\ee
and
\be
\label{Pisdot_new}
\dot{\Pi}_\psi = \left[ N \left( R^{n-2} \psip + \left( \frac{N_r}{N}
\right) \Pis \right) \right]^\prime,
\ee
where dots and primes represent differentiation with respect to $t$ and $R$ respectively.  Preservation of the gauge condition $R=x$ in time  determines the shift algebraically in terms of the lapse, the areal radius and the mass function:
\bea
{\cal M}&:=& \frac{1}{2kG}\left[
R^{n-3}\left(\frac{N_r}{N}\right)^2 + 
\tilde{\alpha}_{(2)}R^{n-5}\left(\frac{N_r}{N}\right)^4 \right]
\label{eq:M EL2}
\eea
This can be solved algebraically and yields, for EGB gravity:
\be
\frac{N_r}{N} =\sqrt{\frac{R^2}{\tilde{\alpha}} \left(  \sqrt{1 +
\frac{2\tilde{\alpha}}{R^2} \frac{2kG {\cal M}}{R^{n-3}}} -1 \right)}.
\label{Nsig_code}
\ee
where we have defined $\tilde{\alpha}:=2(n-4)(n-3) \alpha_{(2)}$ for purposes which will become obvious later.
The sign of the inner square root in the above has been chosen to give the correct GR limit as $\tilde{\alpha}\to0$.

The consistency condition $\dot{\Lambda}=0$ determines the lapse via:
\be
\label{sigma_code}
N^\prime = -\frac{kGN \Pis \psip}{R^{n-3}} \Big/ \left( \sub \left( 1
+ \frac{\tilde{\alpha}}{R^2} \sub^2 \right) \right),
\ee
while the Hamiltonian constraint then takes the form:
\be
\label{Hconstraint_new}
\m^\prime = \frac{1}{2} \left( \frac{\Pis^2}{R^{n-2}} + R^{n-2} \psip^2 \right) + \sub \Pis \psip.
\ee
Using (\ref{Nsig_code}) to replace  $N_r/N$ by ${\cal M}$ in (\ref{Hconstraint_new}) provides a differential equation that can be solved for $\cal M$ and hence $N_r/N$ in terms of the scalar field and its conjugate momenta on each spacial slice.

Given the solution for ${\cal M}$, one can look for apparent horizons by solving (\ref{eq:M EL}) for $(DR)^2=0$. In our coordinates, this becomes simply:
\be
AH:=1- \left(\frac{N_r}{N}\right)^2=0
\label{eq:AH}
\ee
where for ease of reference we refer to $AH$ as the horizon function. 
 For EGB, one can also use (\ref{eq:M EL2}) and the above to obtain:
\be
\label{MAH}
{\cal M}(R_{AH})=\frac{1}{4kG} \left( \tilde{\alpha} R_{AH}^{n-5} + 2R_{AH}^{n-3} \right).
\ee
Note that in 5D the first term is constant so that there is an algebraic lower bound on the black hole mass as the radius of the horizon goes to zero.

Our goal is to solve the time evolution equations (\ref{psidot_new}) and (\ref{Pisdot_new}) for the scalar field and its conjugate momentum, with $N$ and $N_r/N$ on each time slice determined using (\ref{Nsig_code}),(\ref{sigma_code}) and (\ref{Hconstraint_new}) and then use (\ref{MAH}) to look for the formation of an apparent horizon.

The actual time evolution equation as implemented in the code was obtained by expanding the derivative in (\ref{Pisdot_new}) and replacing the derivatives of ${\cal M}'$ and $N'$ using Eqs.(\ref{Nsig_code}) and (\ref{Hconstraint_new}).  This gives:
\begin{align}
\label{Pisdot_code}
& \dot{\Pi}_\psi = N \Bigg\{ \Bigg[ Gk \left( \frac{\Pis^3}{2R^{2n-5}} -
\frac{\psip^2\Pis R}{2} \right) \Big/ \sub + \\ \nonumber
& - \frac{(n-3) \Pis}{2R} \sub - \frac{\tilde{\alpha}(n-5)\Pis}{4R^3} \sub^3
\Bigg]
\Big/ \left( 1 + \frac{\tilde{\alpha}}{R^2} \sub^2 \right) \\ \nonumber
& + (n-2)R^{n-3}\psip + R^{n-2}\psi^{\prime \prime} + \Pis^\prime \sub \Bigg\}.
\end{align}
Note that in five space-time dimensions the last term proportional to $1/R^3$ in the square brackets above vanishes. One might therefore expect behaviour for $n>5$ that is qualitatively different from $n=5$. It is for this reason that it is important to study higher dimensions.  In the present paper we restrict consideration to five and six dimensions.
%%%%%%%%%%%%%%%%%%%%%%%%%%%%%%%
\section{Numerics and Methods}
\label{Numerics and Methods}

The system is evolved using c++ code as follows:

\begin{enumerate}
\itemsep -1mm
\item Initialize the spatial lattice.  We set the lattice spacing to $10^{-5}$ (unless otherwise stated) for the first 100 points near the origin and then slowly increase it to $10^{-2}$ at the 1200$^{th}$ and final lattice point.

\item Set up initial conditions.  We initialized $\Pis$ to zero and $\psi$ to be either a Gaussian $\psi_G$ or hyperbolic tangent $\psi_H$ as follows 
\be
\label{init}
\psi_G = AR^2 \exp\left[-\left(\frac{R-R_0}{B}\right)^2\right]
\, ; \qquad
\psi_H = A \tanh\left[\frac{R-R_0}{B}\right] 
\ee
where $A$, $B$ and $R_0$ are parameters.

\item At $R=0$ set $N=1$, ${\cal M}=0$ and use a subroutine to calculate
$N_r/N$ using equation Eq.(\ref{Nsig_code}).  Integrate $N$ and ${\cal M}$ forward in R using equations Eqs.(\ref{Nsig_code}), (\ref{sigma_code}) and (\ref{Hconstraint_new}). This is done using an RK4 method.  Spatial derivatives are calculated using a central difference routine except at the boundaries where forward and
backward difference are used.

\item Integrate $\psi$ and $\Pis$ forward in time using equations
(\ref{psidot_new}), (\ref{Pisdot_new}) and (\ref{Nsig_code}) employing an RK4 method. Stability is maintained by insisting that the size of the time step, $\Delta
t(t)$, is determined by
\be
\label{stability}
\Delta t(t) < \min_R \left\{ \left(\frac{dR}{dt} \right)^{-1} \Delta R(R)
\right\},
\ee
where $\Delta R(R)$ is the lattice spacing and $\frac{dR}{dt}$ is the maximum
value of either the ingoing or outgoing local speed of light.

\item Monitor the apparent horizon function, $AH:=(DR)^2 =1-\sub^2$.  At
any point where $AH=0$, there is an apparent horizon.  When $AH$ forms a minimum
it signals that an apparent horizon is soon to form, so the time steps are
diminished by a factor of ten.

\item Calculate $\Tm$ and the mass density.

\item Repeat steps 3-6 until the formation of an apparent horizon or until the the field has dispersed. %{\bf Erase the following, it appears later:  In the initial binary search for $A^{*}$, after a pre-determined time limit the code terminated on the assumption that no horizon was going to form. The results were then checked near $A^{*}$ (sub- and super-critical) by direct observation of the evolution of the field at late times.}

\end{enumerate}

For comparison purposes it was important that the code could simulate collapse
without the GB term, ie in the GR case.  %{\bf While this can be achieved analytically by taking the limit as $\tilde{\alpha} \to 0$ in the equations of motion,  it is}
It is not
possible to take this limit when numerically calculating $N_{r}/N$ using
Eq.(\ref{Nsig_code}) so an if statement was added to the routine
which calculates $N_r/N$ in order to return $N_r/N=\sqrt{2kG{\cal M}/R^{n-3}}$ when
$\tilde{\alpha} = 0$.  

When $\tilde{\alpha}$ is not zero a problem arises in the calculation of
$N_r/N$ when $4\tilde{\alpha} kG{\cal M}/R^{n-1}$ is sufficiently less than one.  When this term is added to unity in the inner square root in (\ref{Nsig_code}), digits are lost and thus double precision can not be claimed.  For this reason a 16$^{th}$ order Taylor expansion of the inner square root in
 Eq.(\ref{Nsig_code}) was used in the case that $2\tilde{\alpha} 2kG{\cal M}/R^{n-1}<0.1$ Quadruple precision allowed for the investigation of overflow
and underflow, as well as subtraction and addition round off errors.
%Quadruple precision allowed for the investigation of overflow nd underflow, as well as subtraction and addition round off errors. To urther reduce the effects of round off error and lost digits the code was made capable of quad (34 digit) precision.  
 
The code was 
capable of parallel processing,  and many
simulations were run on eight or more processors using the WestGrid and SHARCNET computing clusters.  When generating data for mass
and $\Tm$ scaling plots the speed up was linear with the number of processors
used, whereas for the binary search used to find critical values the speed
up was logarithmic.

\begin{figure}[ht!]
\centering
\subfigure[5D]{
\includegraphics[width=0.4\linewidth]{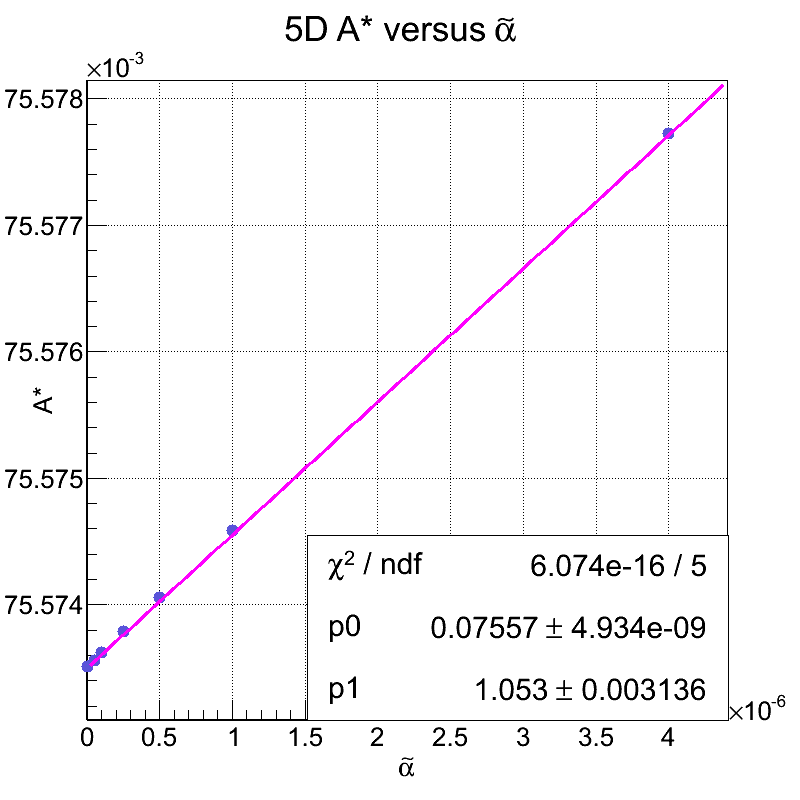}
\label{Astar5}
}
\hspace{0.25in}
\subfigure[6D]{
\includegraphics[width=0.4\linewidth]{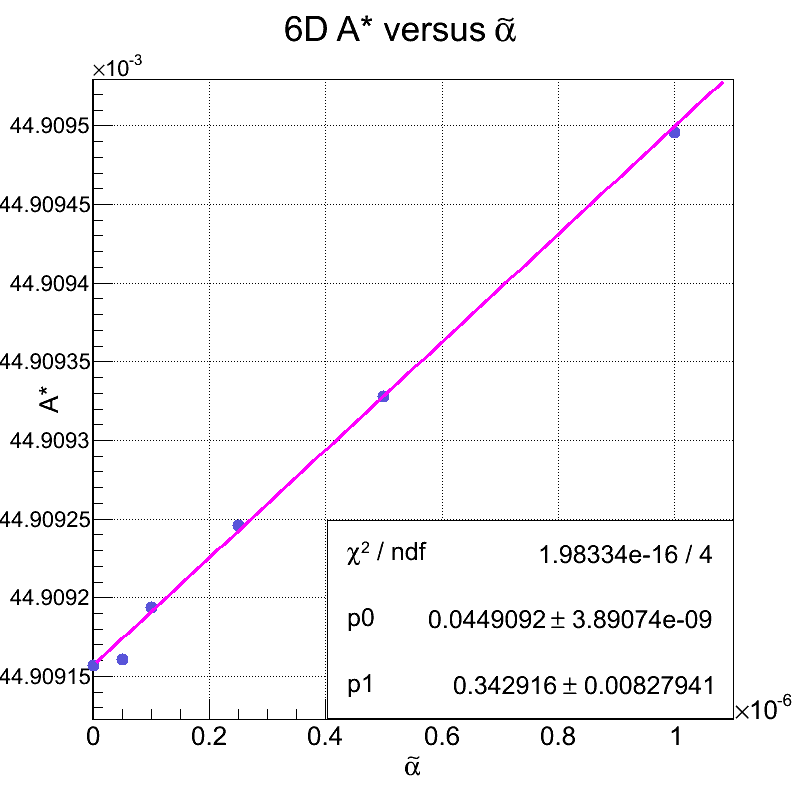}
\label{Astar6}
}
\caption{$A^*$ as a function of $\tilde{\alpha}$}
\label{Astar}
\end{figure}

The critical value of a parameter in the initial conditions is defined as the
value of that parameter for which a black hole just barely forms.  We first
performed a binary search to find the critical value of $A$ in Eq.(\ref{init}).  $\psi(t, R=0)$ and ${\cal M}$ were then checked  at late times confirm that they blew up for $A$ slightly bigger than $A^*$ and remained finite for $A$ slightly smaller than $A^*$.  We were able to get consistent results to 12 significant figures.  The $A^*$ values for different values of $\tilde{\alpha}$ can be seen in figure \ref{Astar}.  Interestingly the points are very well fit to straight lines.  The above procedure, of course, also gives $B^*$ and $R_0^*$, which could also be varied.  Using our values for $A^*$ we calculated the wave function at the origin as a function of PG time and used these plots to find the period of oscillation near criticality, as a function of the $\tilde{\alpha}$.   

We simulated matter bounce and dispersal for 280 simulations (the number 280 chosen to optimize graph resolution and computing time) with $A<A^*$ and recorded $\Tm$ for each simulation.  Plotting $\Tm$ as a function of $A^*-A$ with a log-log scale gives the $\Tm$ scaling plot.  Similarly we simluated collapse for 280 simulations with $A>A^*$ and recorded  the radius of the initial apparent horizon, $R_{AH}$.  This procedure was repeated in five and six dimensions checking for scaling with both
the $A$ and $B$ parameters in both the gaussian and tanh initial data of equation \ref{init} to check for universality.  Using gaussian initial data radius and $\Tm$ scaling plots were created for $\tilde{\alpha} = 10^{-8}, 10^{-7}, 5\times10^{-7}, 10^{-6}$ as well as the GR case $\tilde{\alpha}=0$ to investigate the effects of the GB terms on the critical exponent, period and the existence of mass gap.

% Since $\tilde{\alpha}$ is the only dimensionful parameter in the action, it should set the scale for the dynamical behaviour. Thus in any given space-time dimension one expects qualitatively similar behaviour near criticality for all values of $\tilde{\alpha}$. The only thing that changes is the difficulty of the numerics. If $\tilde{\alpha}$ is very small then it is difficult to get sufficient spatial resolution to explore the GB region. If $\tilde{\alpha}$ is too large, then the numerics appears to be problematic quite far from the critical region.  We therefore give particular attention to one specific value of $\tilde{\alpha}$ in each dimension}, chosen to optimize the numerics. In five dimensions, this is $\tilde{\alpha}=10^{-6}$, whereas in six dimensions $\tilde{\alpha}=10^{-5}$. We have also verified that other values of $\tilde{\alpha}$ do produce qualitatively similar scaling plots over the range that we were able to reliably calculate them.

%The period of oscillation of $f$ was also calculated in both regions.  {\bf [Got to do this still.]}

\section{Results}
\label{Results}
\subsection{Scalar Field Oscillations}
\begin{figure}[ht!]
\centering
\subfigure[5D,$\psi(0,t)$ near criticality, GR]{
\includegraphics[width=0.4\linewidth]{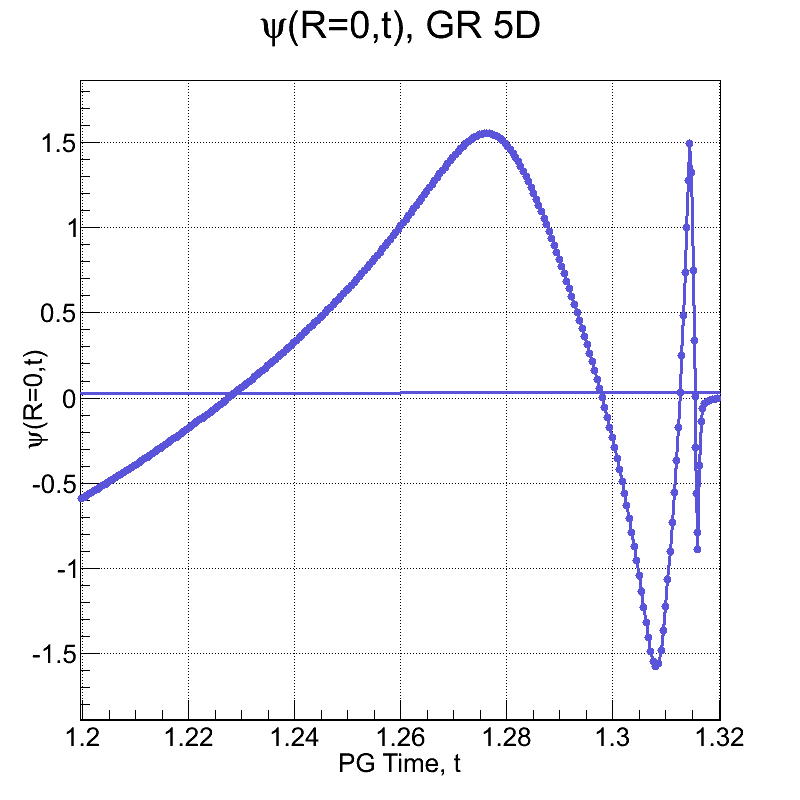}
\label{5Dscalar1_gr}
}
\hspace{0.25in}
\subfigure[5D,$\psi(0,t)$ near criticality, $\tilde{\alpha}=10^{-6}$]{
\includegraphics[width=0.4\linewidth]{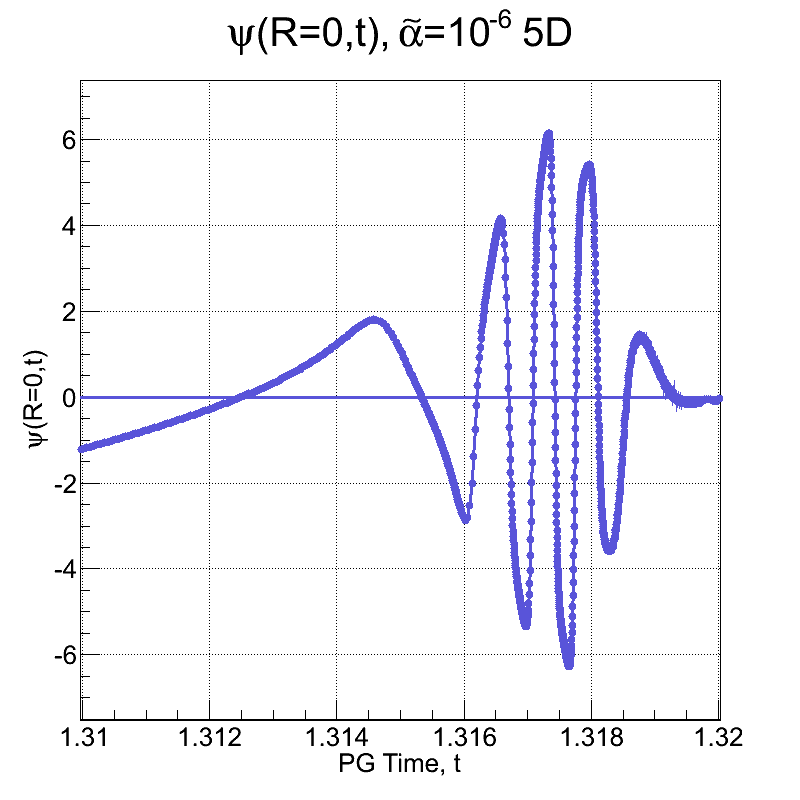}
\label{5Dscalar1a}
}
\subfigure[5D, Period of $\psi(0,t)$ near criticality, $\tilde{\alpha}=10^{-7}$, showing convergence]{
\includegraphics[width=0.4\linewidth]{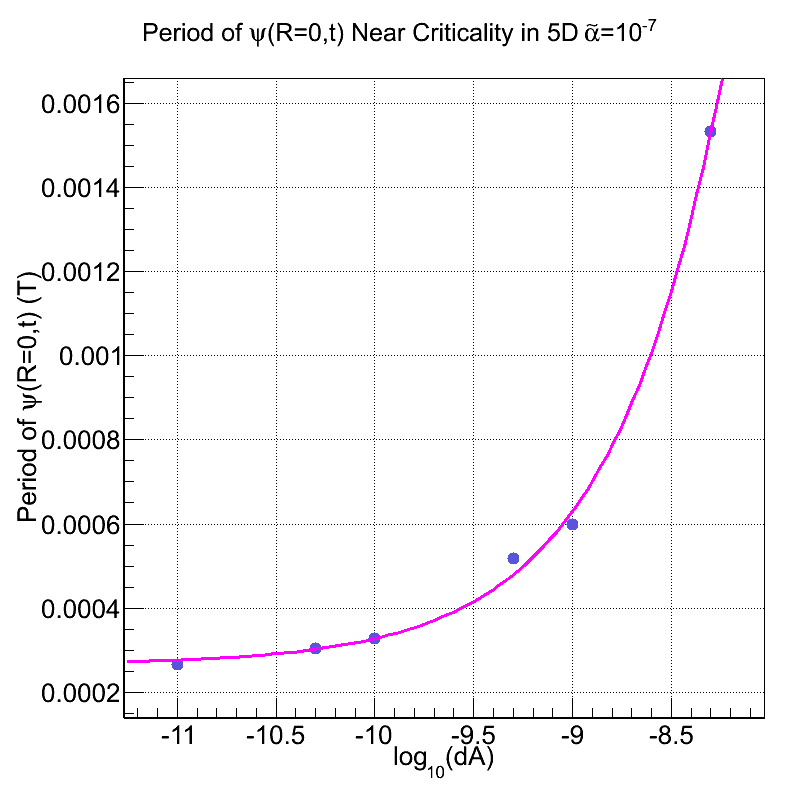}
\label{5Dscalar1b}
}
\hspace{0.25in}
\subfigure[Period of $\psi(0,t)$ in 5D near criticality as a function of GB parameter]{
\includegraphics[width=0.4\linewidth]{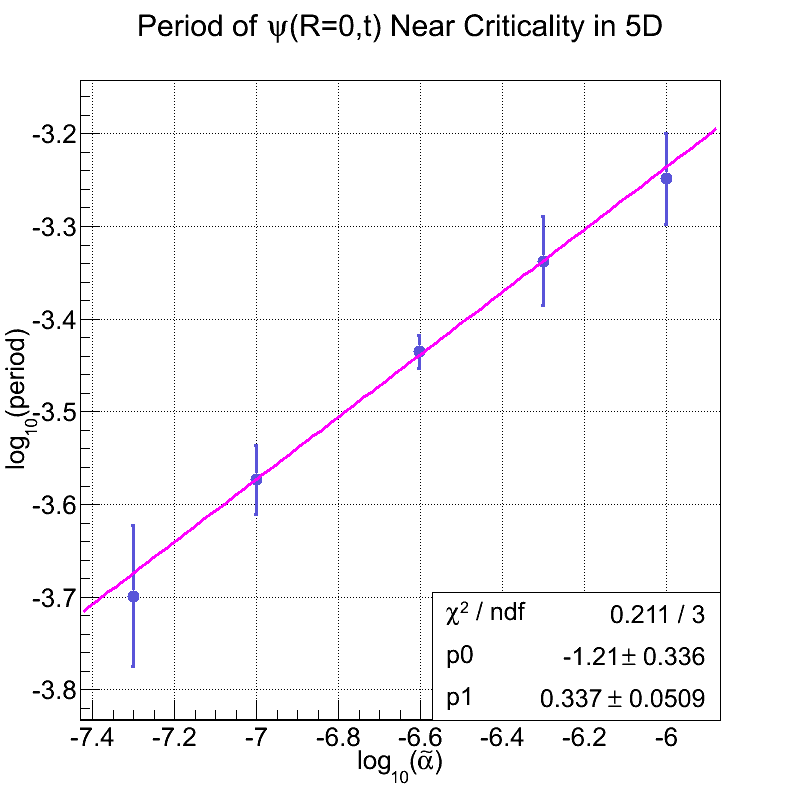}
\label{5Dscalar1c}
}
%\hspace{0.25in}
%\subfigure[5D, scalar oscillation period as function of GB parameter]{
%\includegraphics[width=0.4\linewidth]{xxx}
%\label{5Dscalar1b} }
\caption{Scalar field oscillations}
\label{5Dscalar}
\end{figure}

\begin{figure}[ht!]
\centering
\subfigure[6D,$\psi(0,t)$ near criticality, GR.]{
\includegraphics[width=0.4\linewidth]{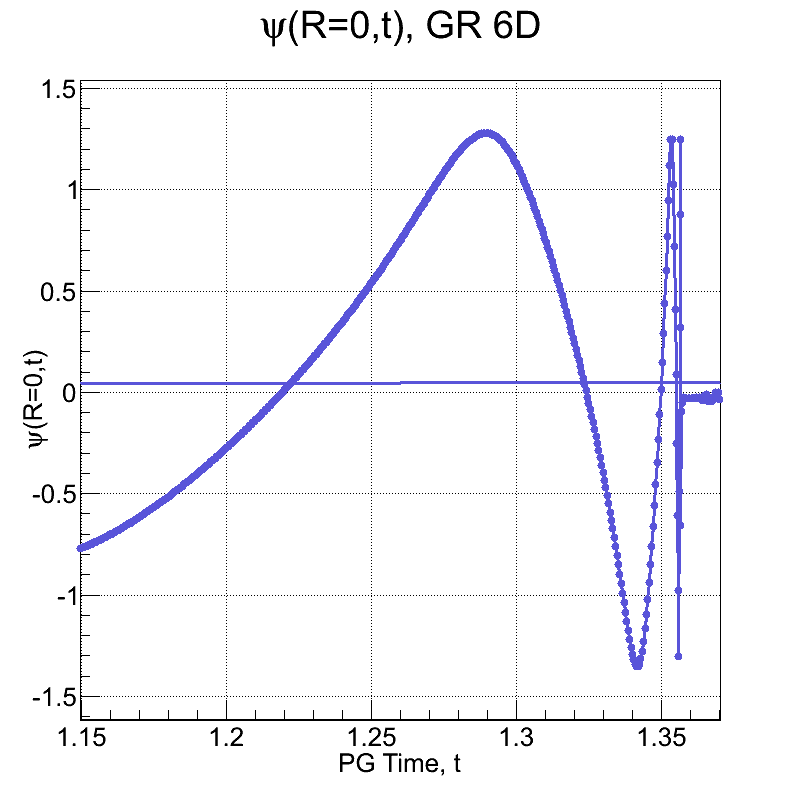}
\label{6DscalarGR}
}
\hspace{0.25in}
\subfigure[6D,$\psi(0,t)$ near criticality, $\tilde{\alpha}=10^{-6}$.]{
\includegraphics[width=0.4\linewidth]{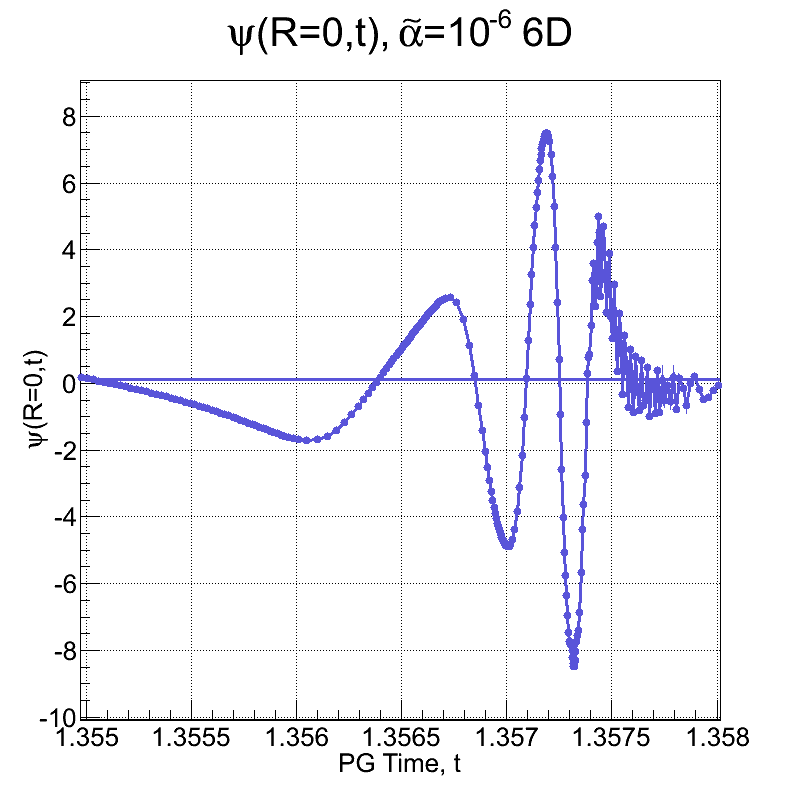}
\label{6Dscalar1a}
}
\subfigure[6D, period of $\psi(0,t)$ near criticality, $\tilde{\alpha}=10^{-7}$, showing convergence.]{
\includegraphics[width=0.4\linewidth]{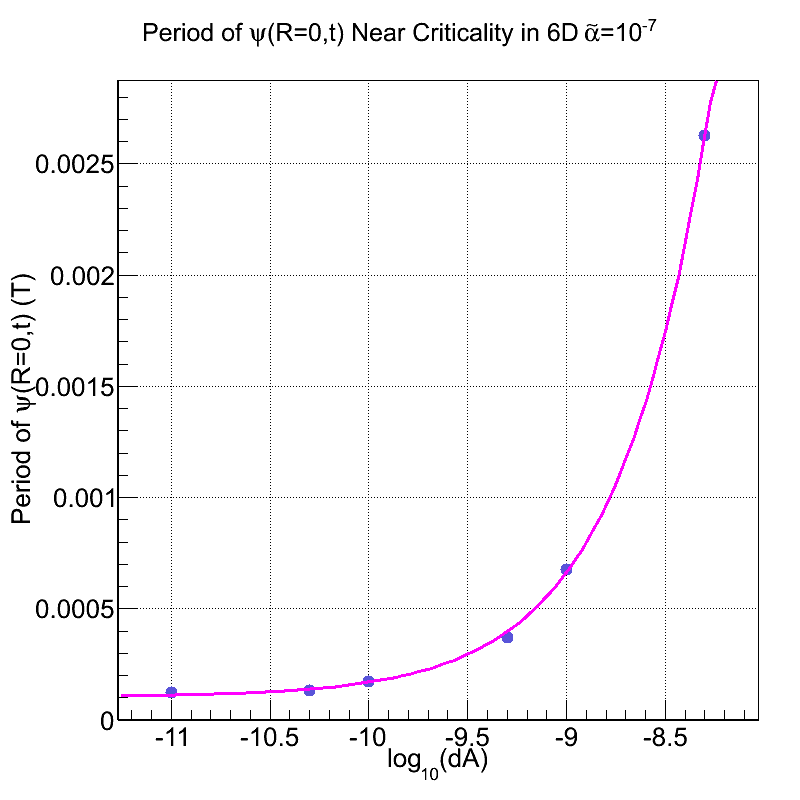}
\label{6Dscalar1b}
}
\hspace{0.25in}
\subfigure[Period of $\psi(0,t)$ in 6D near criticality as a function of GB parameter]{
\includegraphics[width=0.4\linewidth]{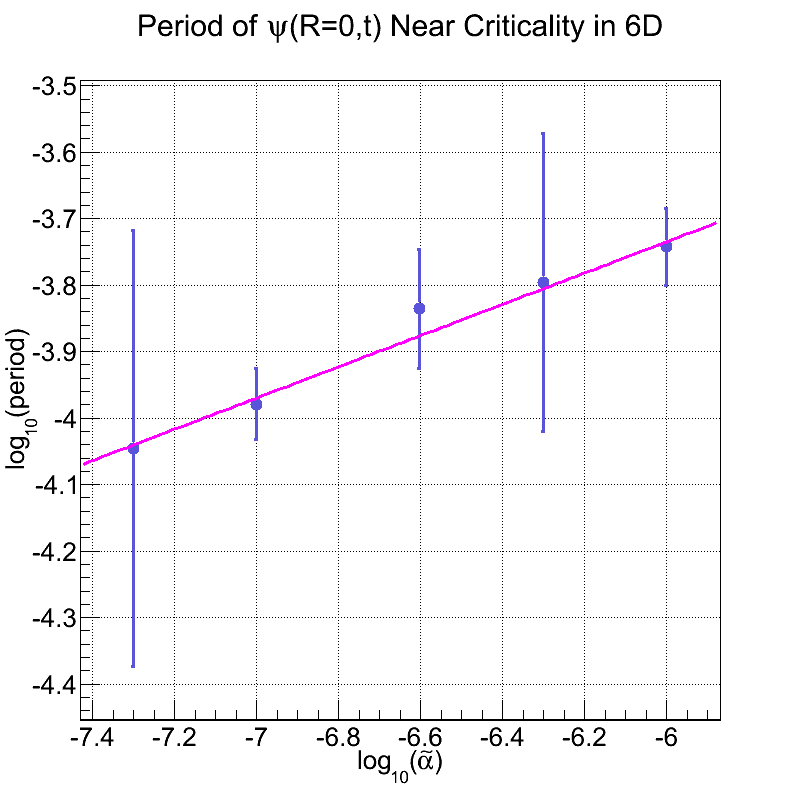}
\label{6Dscalar1c}
}
%\hspace{0.25in}
%\subfigure[6D, scalar oscillation period as a function of GB parameter]{
%\includegraphics[width=0.4\linewidth]{xxx}
%\label{6Dscalar1b} }
\caption{Scalar field oscillations}
\label{6Dscalar}
\end{figure}

In general relativity the discrete self-similarity of the critical solution results in oscillations of the scalar field at the origin with ever decreasing period. The presence of the dimensionful Gauss Bonnet parameter breaks the scale invariance and the discrete self-similarity  \cite{Golod2012}. The scalar field oscillations at the origin near criticality approach a constant period that depends on the value of the GB parameter. Since it was difficult to get close enough to criticality to guarantee that the period had converged, we plotted the values as a function of $\log(dA)$, where $dA\equiv |A-A^*|$. As seen in  Figs.\ref{5Dscalar1b},\ref{6Dscalar1b} the convergence was exponential and we used a best fit to determine the value of the period $T$ and its corresponding error for each value of $\tilde{\alpha}$. The results are shown for 5 and 6 dimensions in Figs.\ref{5Dscalar1c},\ref{6Dscalar1c}. Our results are qualitatively similar to those in  \cite{Golod2012}, namely
\bea
T_{(n)}&\propto& \tilde{\alpha}^\beta_{(n)} \label{eq:lambda}
\eea
Our exponents in five and six dimensions are:
\bea
\beta_{(5)} &=& 0.34\pm0.05\\
\beta_{(6)} &=& 0.24\pm0.08
\eea
These both differ from the value of approximately 1/2 obtained in 5D by Golod and Piran  \cite{Golod2012}, who  argued that $\beta$ is one divided by the scaling dimension of the GB coupling coefficient. Intriguingly our results suggest a relationship of 
\be
\beta_{(n)} = 1/(n-2)
\ee
%which, oddly enough, is one divided by the dimensionality of the GR coupling constant.
Note that the 6D plots show oscillations at late times which are likely due to the build up of numerical error.

We note also that the range of $\tilde{\alpha}$ that we considered was between $5 \times 10^{-8}$ and $10^{-6}$, which is outside the range $4\times10^{-6}$ to $4\times10^{-4}$ considered by  \cite{Golod2012}, which may explain the discrepancy. We were restricted to smaller values of the GB parameter because our PG co-ordinate code did not allow us to get close enough to criticality for large values of $\tilde{\alpha}$ in order to reliably measure the period of the scalar field.
%%%%%%%%%%%%%%%%%%%%%%%%%%%%%%%%%%%%%%%%%%%%%%%%%%%%%%%%%%

%%%%%%%%%%%%%%%%%%%%%%%%%%%%%%%%%%%%%%%%%%%%%%%%%%%%%%%%

%%%%%%%%%%%%%%%%%%%%%%%%%%%%%%%%%%%%%%%%%%%%%%%%%%%%%%%%%%
\subsection{Critical Exponents}
In GR there exist universal scaling relations whose properties are determined in part by the critical solution. We now present two different sets of scaling plots in the GB case. The first is the value of the logarithm of the apparent horizon radius $R_{AH}$ on formation as a function of $\log(dA)$ as the critical parameter is approached from above (i.e. supercritical). The second is the log of $\Tm$ as a function of $\log(dA)$. We find as expected that if we are far enough from criticality that the curvatures stay small  and the apparent horizon radius is large compared to the GB scale, we reproduce approximately the GR results: the curves are universal, with slope approximately equal to the GR critical exponent. The $\Tm$ in this region are approximately straight lines with a small oscillation superimposed, whereas the radius plots show the large amplitude cusps observed in  \cite{Ziprick2009b,Taves2011}.

\begin{figure}[ht!]
\centering
\subfigure[5D $\tilde{\alpha}=5\times10^{-7}$, amplitude and width separately]{
\includegraphics[width=0.4\linewidth]{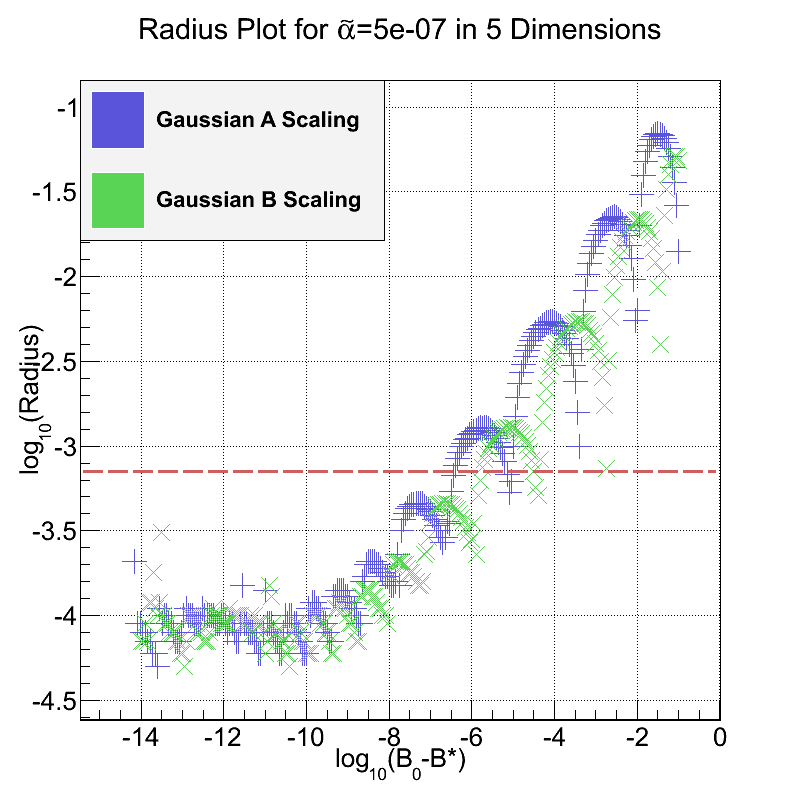}
\label{5DUniverality1}
}
\hspace{0.25in}
\subfigure[5D $\tilde{\alpha}=5\times10^{-7}$, amplitude and width shifted to lie on top of each other]{
\includegraphics[width=0.4\linewidth]{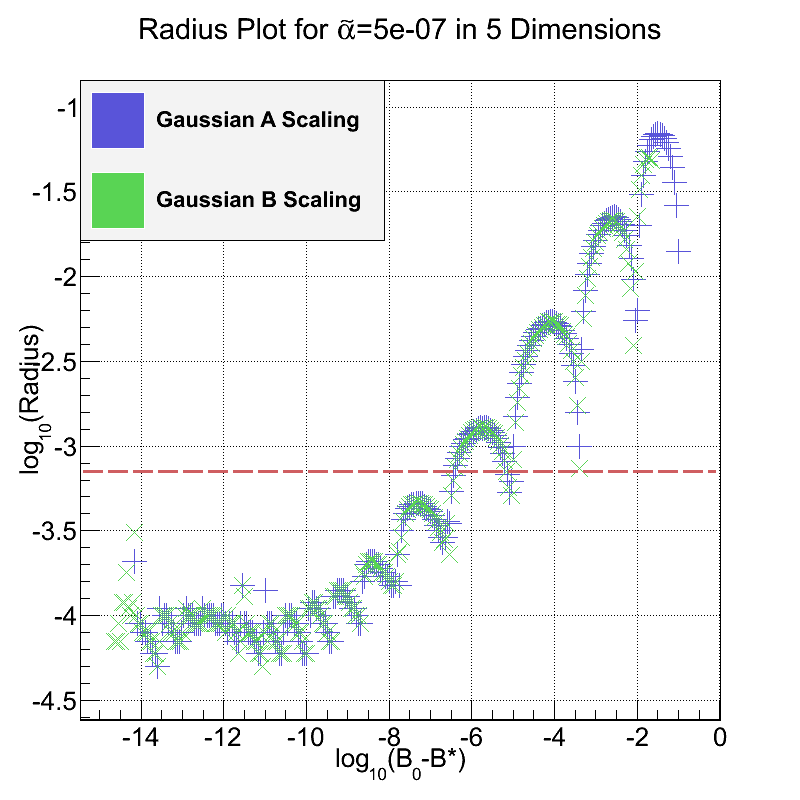}
\label{5DUniversality2}
}
\subfigure[5D $\tilde{\alpha}=5\times10^{-7}$, Radius Plots Superimposed]{
\includegraphics[width=0.4\linewidth]{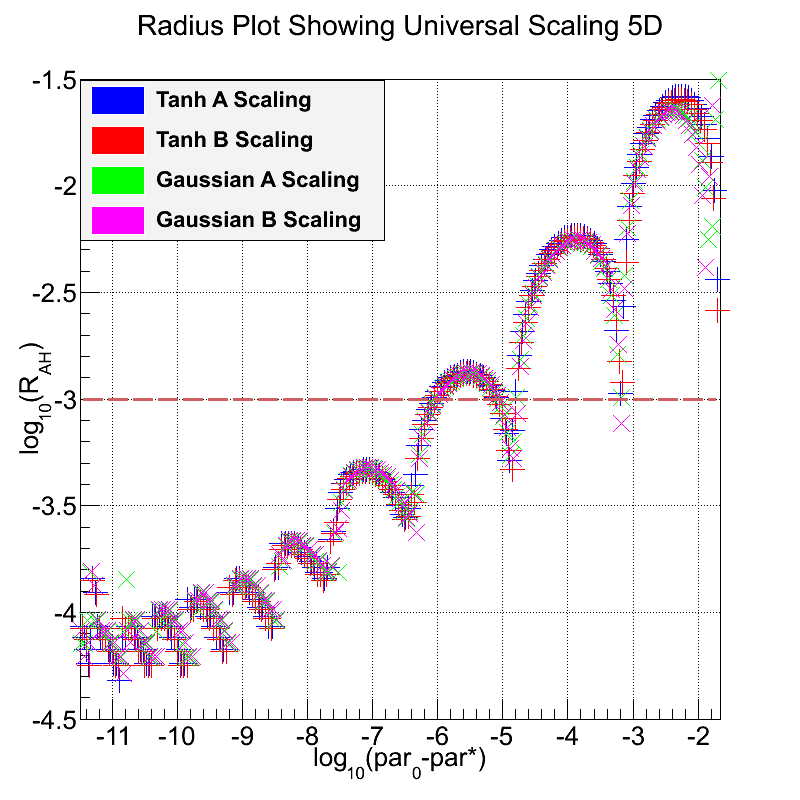}
\label{5DUniversality2b}
}
\hspace{0.25in}
\subfigure[5D $\tilde{\alpha}=5\times10^{-7}$, $\Tm$ Plots Superimposed]{
\includegraphics[width=0.4\linewidth]{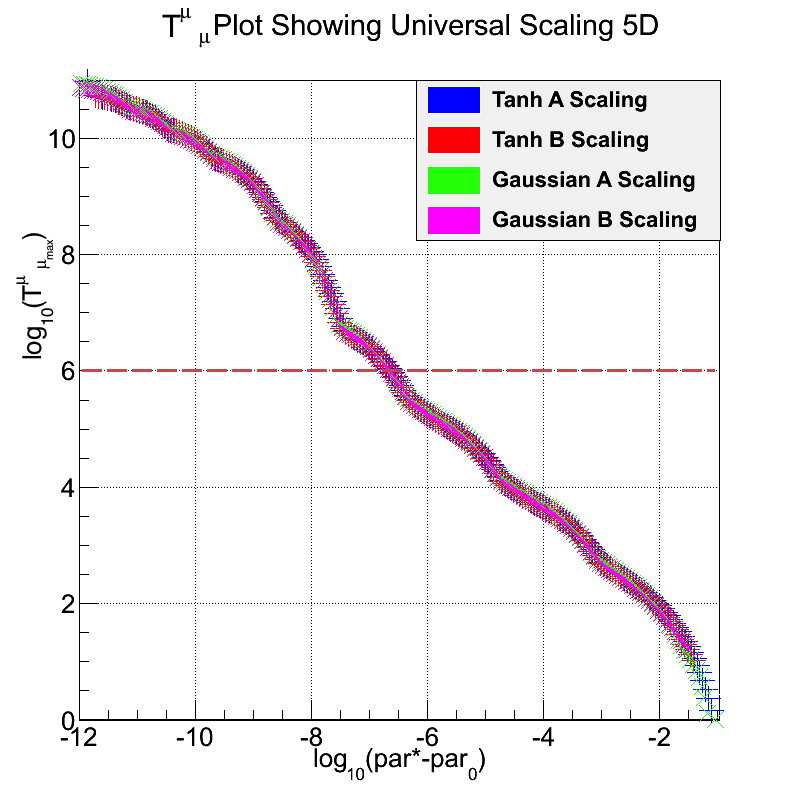}
\label{5DUniverality3}
}
\subfigure[6D, $\tilde{\alpha}=10^{-5}$, Radius Plots Superimposed]{
\includegraphics[width=0.4\linewidth]{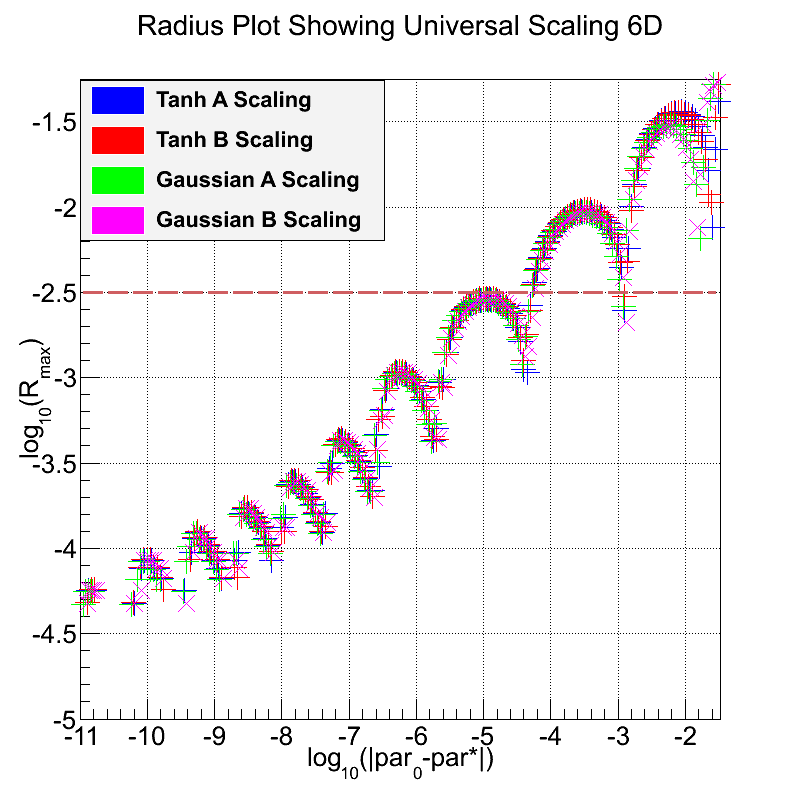}
\label{6DradiusUniversal}
}
\hspace{0.25in}
\subfigure[6D, $\tilde{\alpha}=10^{-5}$, $\Tm$ Plots Superimposed]{
\includegraphics[width=0.4\linewidth]{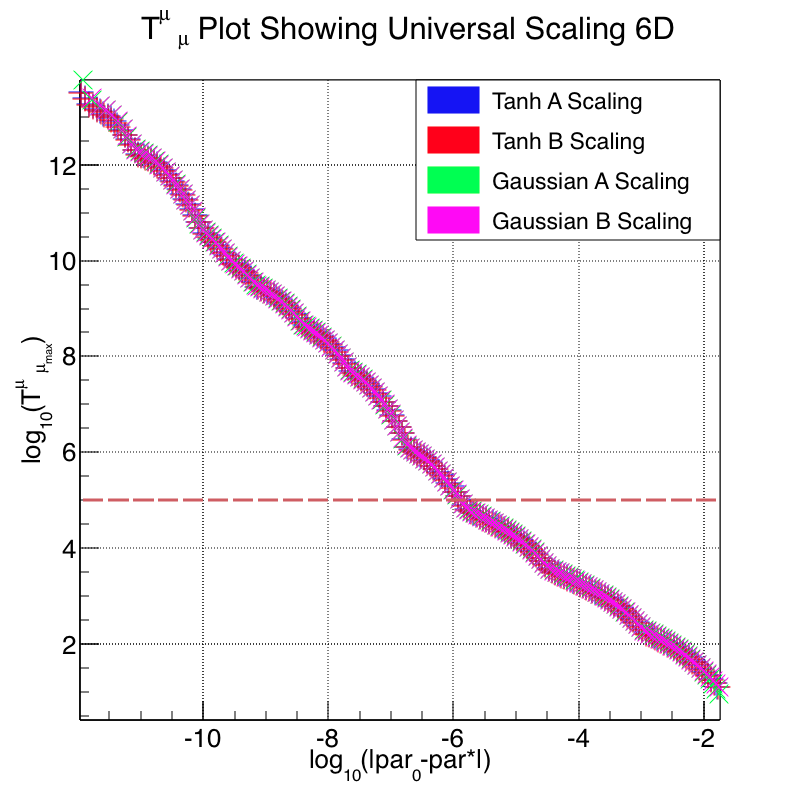}
\label{6DricciUniversal}
}
\caption{Universality in 5 and 6D}
\label{Universality}
\end{figure}

As the critical parameter is approached and we enter into the GB region things change. In the case of the $\Tm$ plots, the GB region occurs when $\Tm \tilde{\alpha} > 1$, whereas for the radius plots it can be defined by the simpler relation $R_{AH}< \sqrt{\tilde{\alpha}}$. The boundary between the two regions is indicated in all the scaling plots by a horizontal dashed line.
  
\begin{figure}[ht!]
\centering
\subfigure[GR, slope$=0.413$]{
\includegraphics[width=0.4\linewidth]{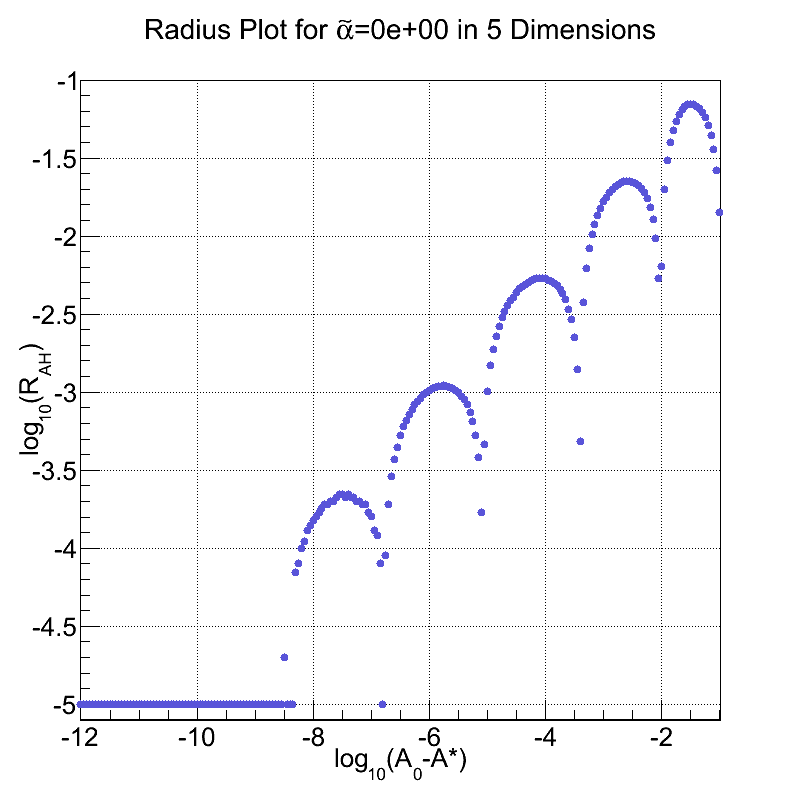}
\label{5DRadiusGR}
}
\hspace{0.25in}
\subfigure[$\tilde{\alpha}=10^{-8}$]{
\includegraphics[width=0.4\linewidth]{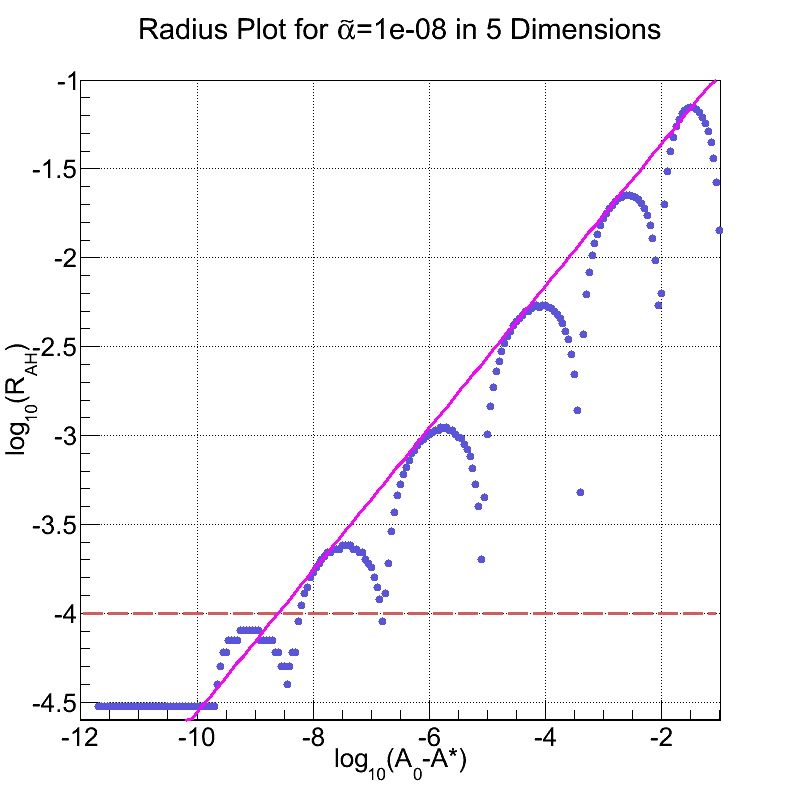}
\label{5DRadius-8}
}
\subfigure[$\tilde{\alpha}=10^{-7}$]{
\includegraphics[width=0.4\linewidth]{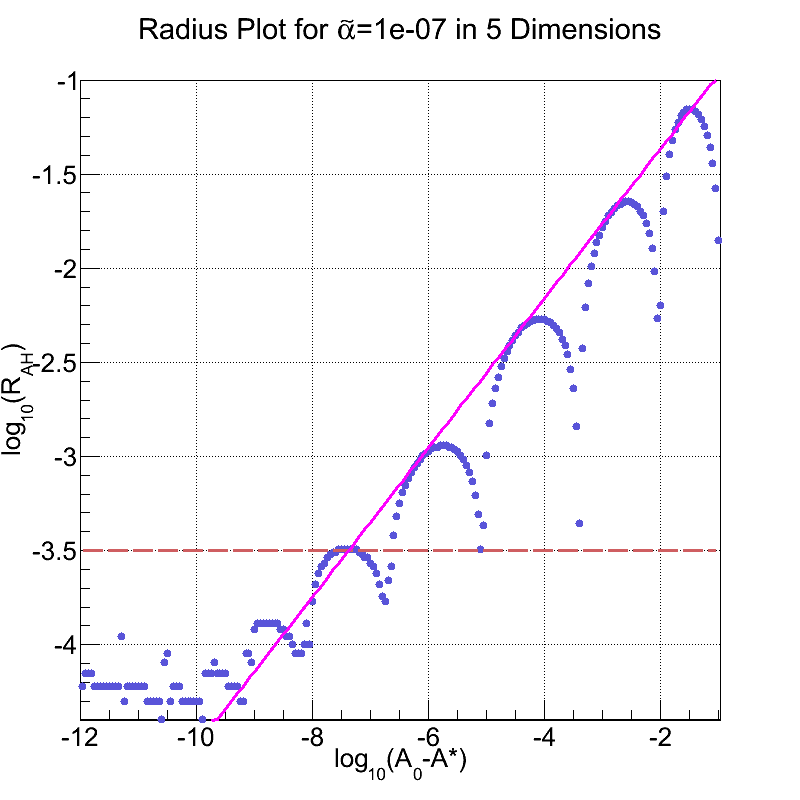}
\label{5DRadius-7}
}
\hspace{0.25in}
\subfigure[$\tilde{\alpha}=5\times10^{-7}$]{
\includegraphics[width=0.4\linewidth]{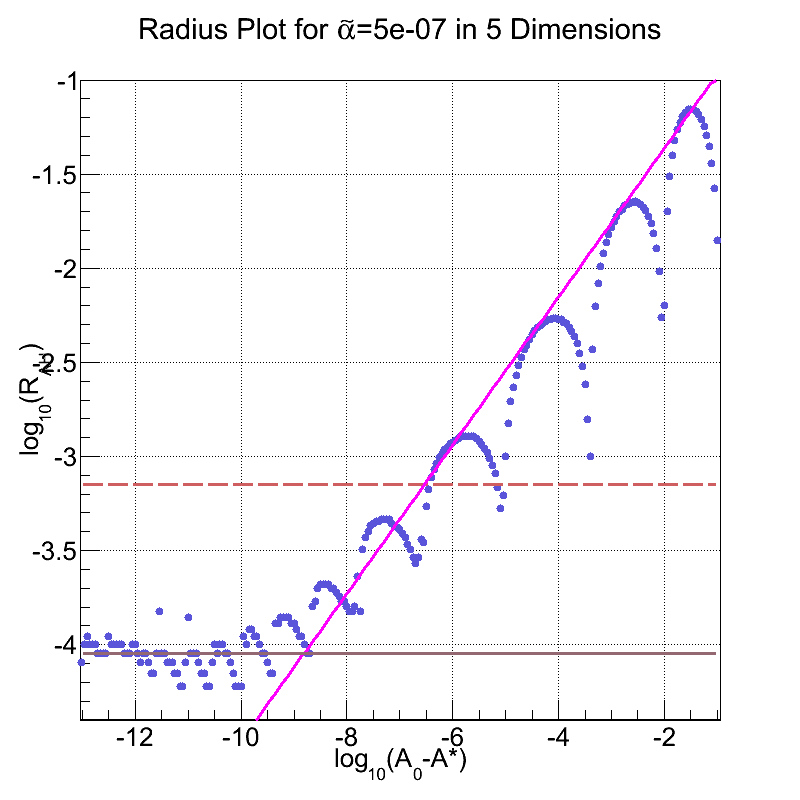}
\label{5DRadius5e-7}
}
\hspace{0.25in}
\subfigure[$\tilde{\alpha}=10^{-6}$]{
\includegraphics[width=0.4\linewidth]{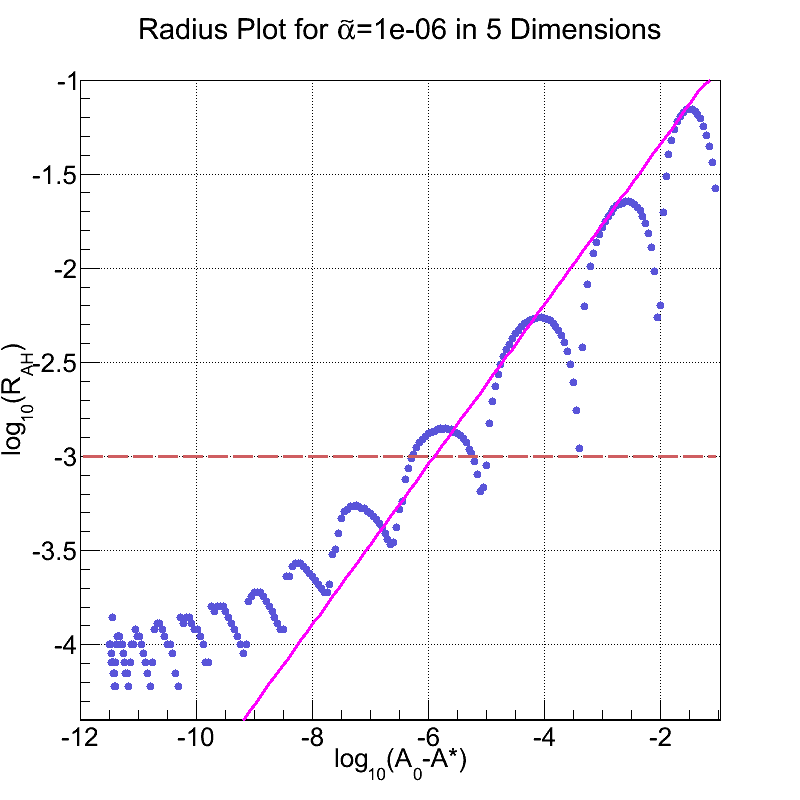}
\label{5DRadius-6}
}
\caption{Radius Scaling Plots - 5D. The lines represent the best-fit tangents to the curves in their respective regimes.}
\label{Radius5D}
\end{figure}

\begin{figure}[ht!]
\centering
\subfigure[GR, slope$=0.43$]{
\includegraphics[width=0.4\linewidth]{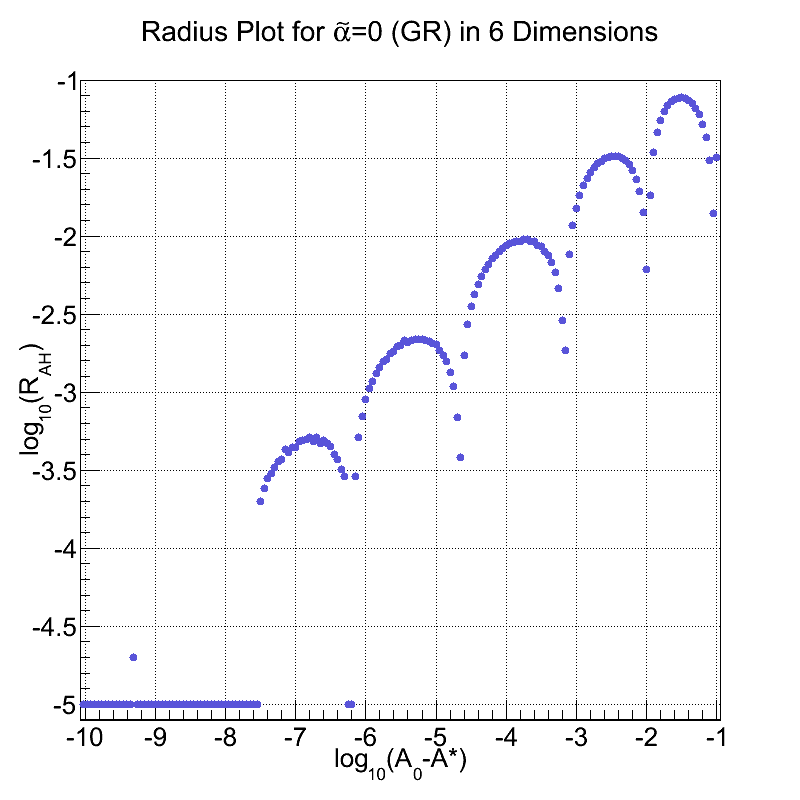}
\label{6DRadiusGR}
}
\hspace{0.25in}
\subfigure[$\tilde{\alpha}=10^{-7}$]{
\includegraphics[width=0.4\linewidth]{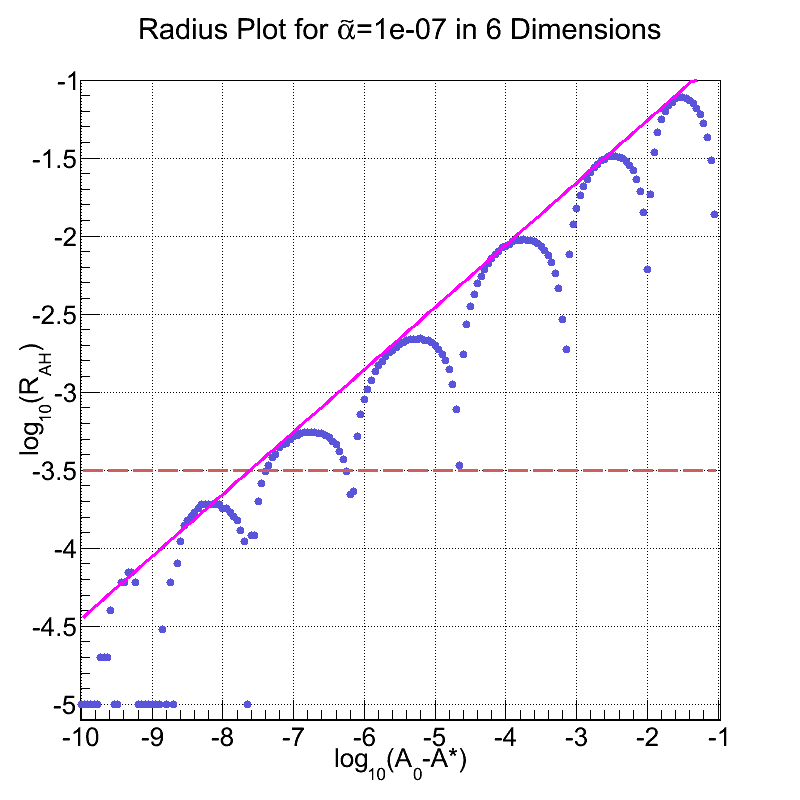}
\label{6DRadius-7}
}
\subfigure[$\tilde{\alpha}=5\times10^{-7}$]{
\includegraphics[width=0.4\linewidth]{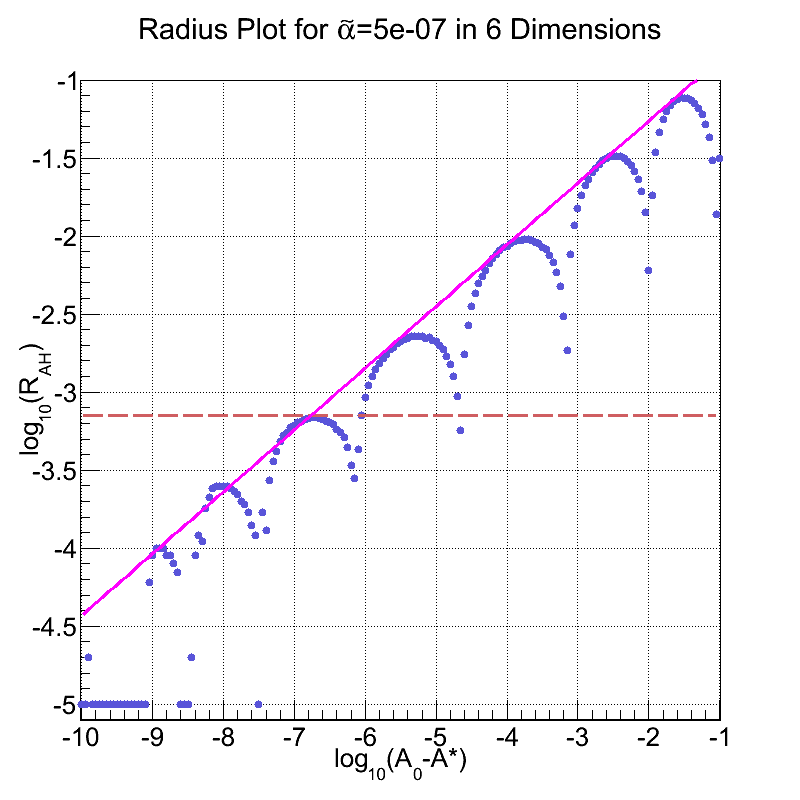}
\label{6DRadius5e-7}
}
\hspace{0.25in}
\subfigure[$\tilde{\alpha}=10^{-6}$]{
\includegraphics[width=0.4\linewidth]{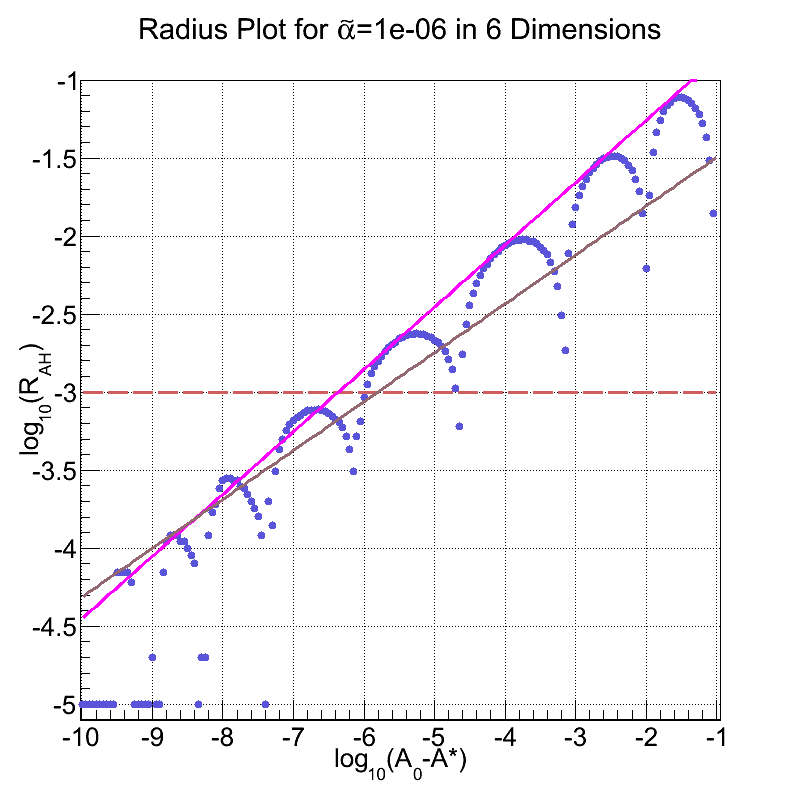}
\label{6DRadius-6}
}
\subfigure[$\tilde{\alpha}=10^{-5}$]{
\includegraphics[width=0.4\linewidth]{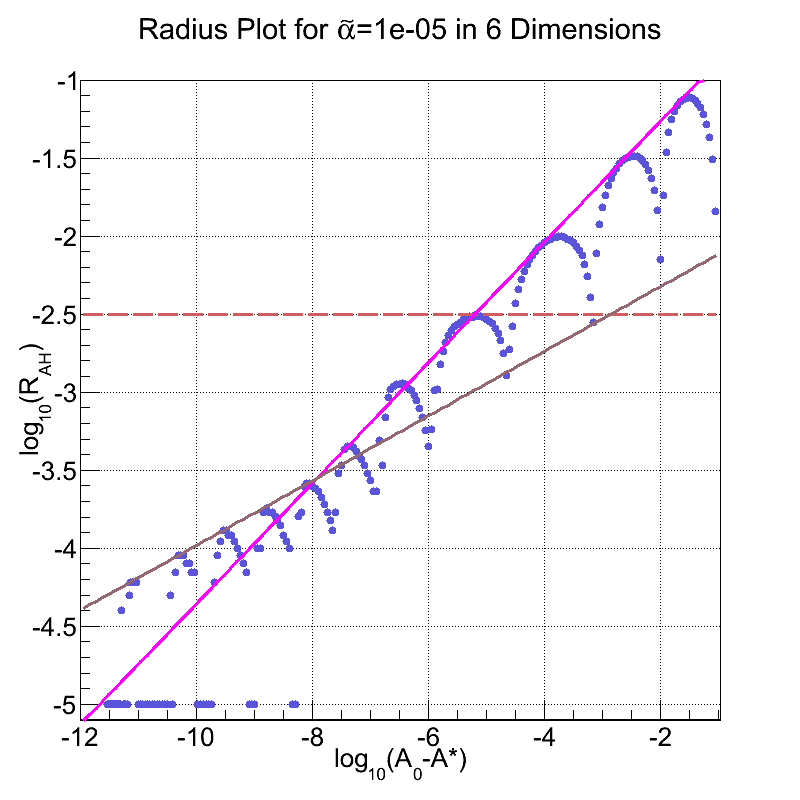}
\label{6DRadius-5}
}
\hspace{0.25in}
\subfigure[$\tilde{\alpha}=10^{-4}$]{
\includegraphics[width=0.4\linewidth]{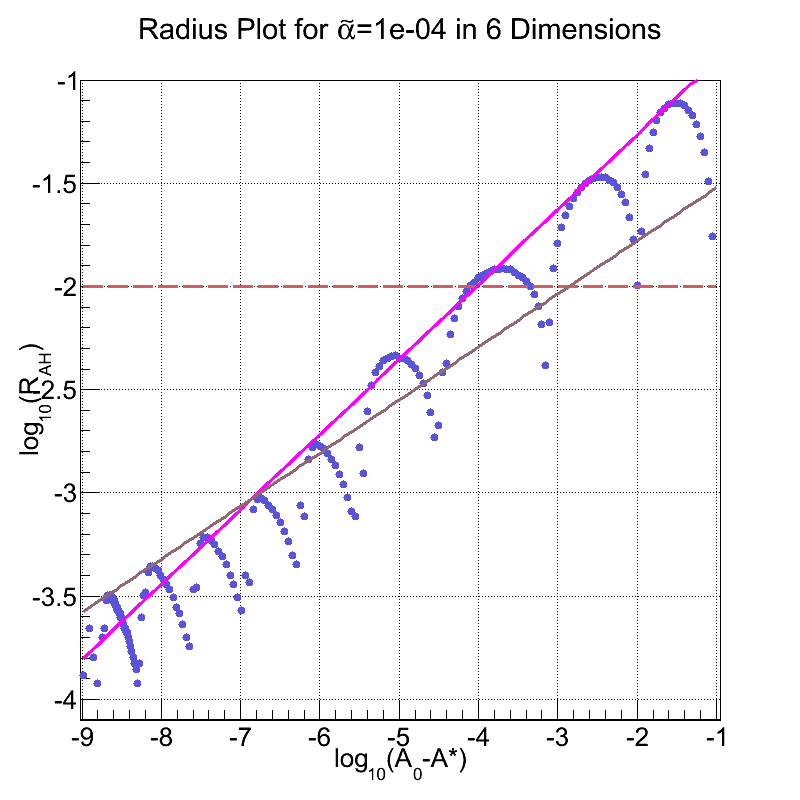}
\label{6DRadius-4}
}
\caption{Radius Scaling Plots - 6D. The lines represent the best-fit tangents to the curves in their respective regimes.}
\label{Radius6D}
\end{figure}

The radius plots Figs.(\ref{Radius5D}) and (\ref{Radius6D}) continue to exhibit cusps, but with a decreased period and slope. The $\Tm$ plots are also similar in the GB region to the GR region in that they are approximately straight lines with oscillations superimposed. However, the slope changes quite suddenly when the transition from GR to GB is made. The first important point is that the scaling plots are universal even in the GB region. This is illustrated for both 5D and 6D in Fig.(\ref{Universality}). There are qualitative differences in the scaling plots between 5D and 6D so we will now discuss the two cases separately.

\begin{figure}[ht!]
\centering
\subfigure[GR, slope$=0.826$]{
\includegraphics[width=0.4\linewidth]{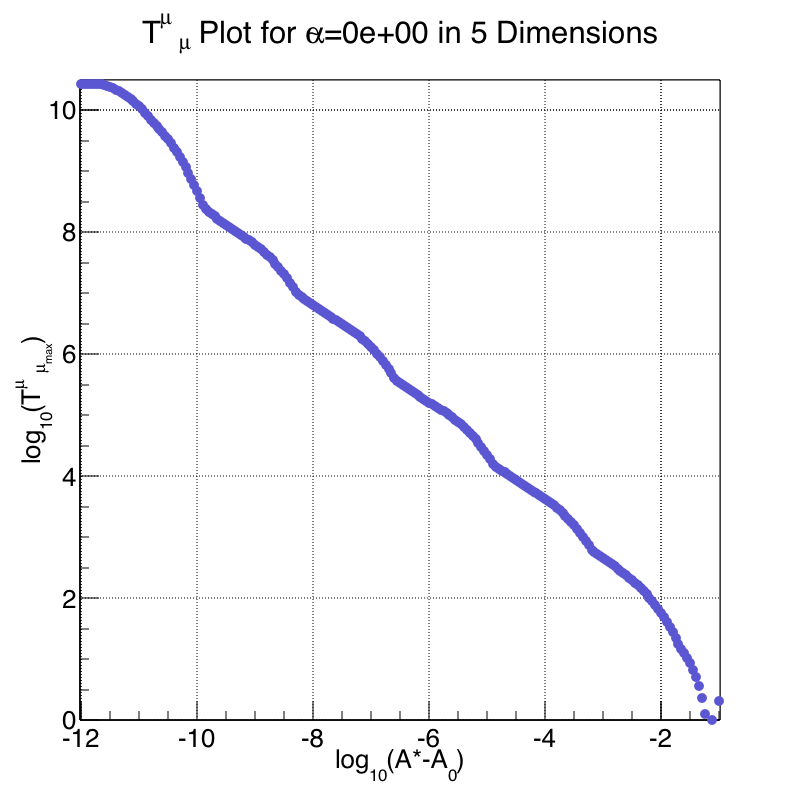}
\label{5DRicciGR}
}
\hspace{0.25in}
\subfigure[$\tilde{\alpha}=10^{-8}$]{
\includegraphics[width=0.4\linewidth]{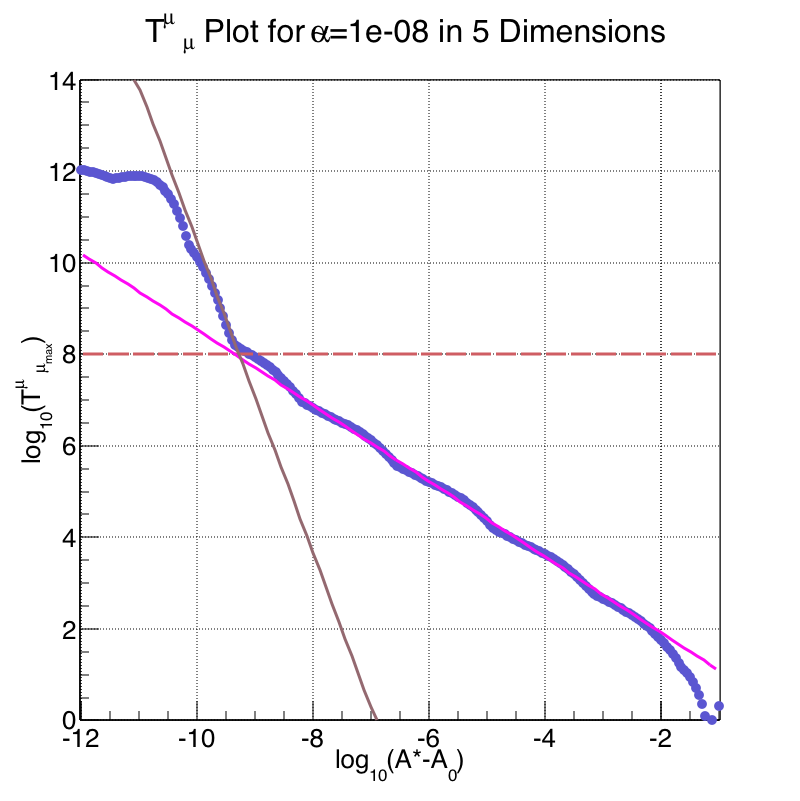}
\label{5DRicci-8}
}
\subfigure[$\tilde{\alpha}=10^{-7}$]{
\includegraphics[width=0.4\linewidth]{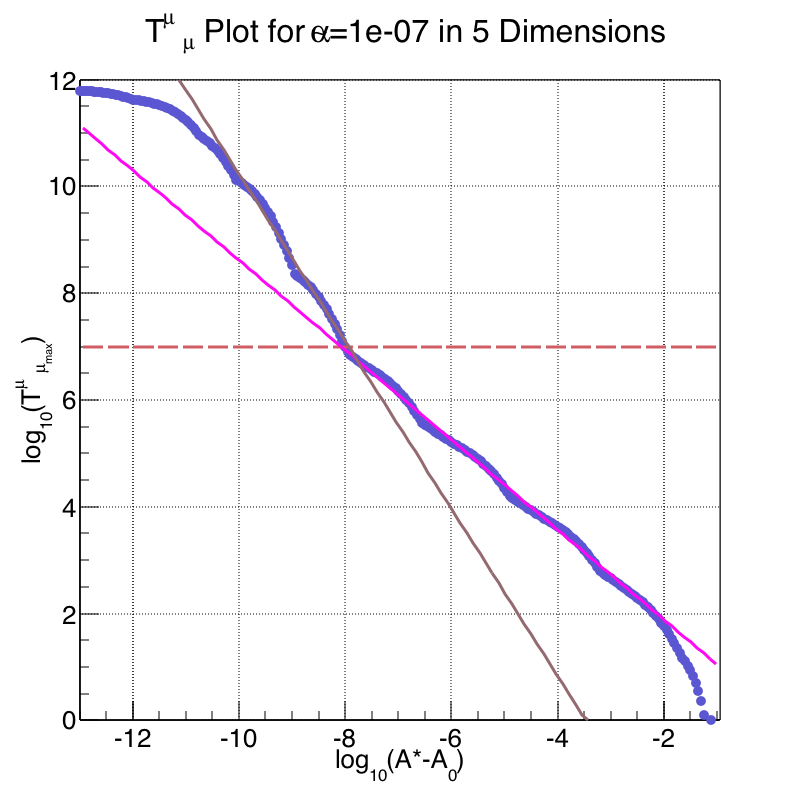}
\label{5DRicci-7}
}
\hspace{0.25in}
\subfigure[$\tilde{\alpha}=5\times10^{-7}$]{
\includegraphics[width=0.4\linewidth]{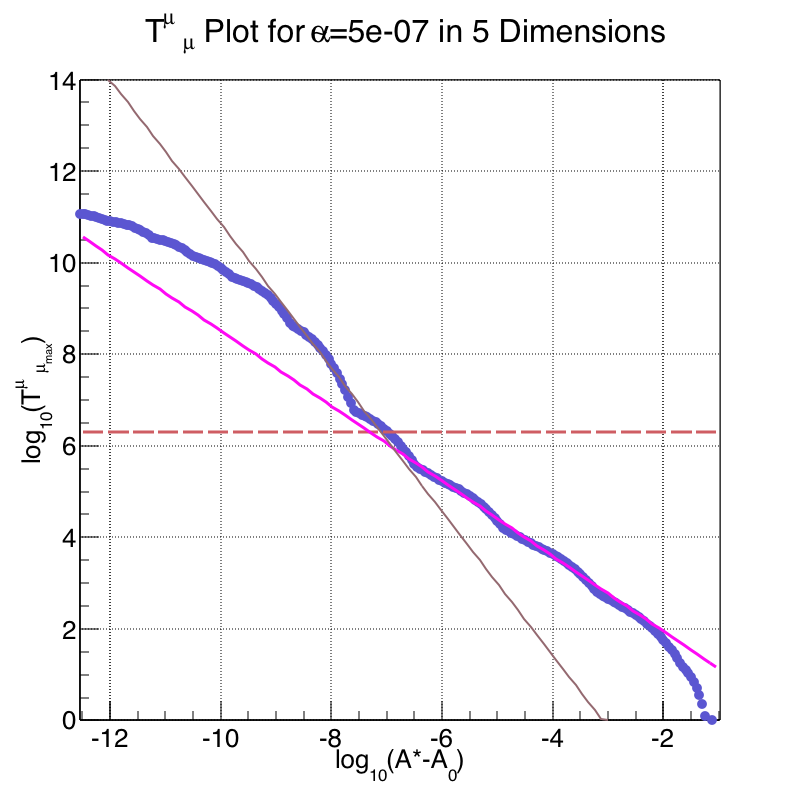}
\label{5DRicci5e-7}
}
\subfigure[$\tilde{\alpha}=10^{-6}$]{
\includegraphics[width=0.4\linewidth]{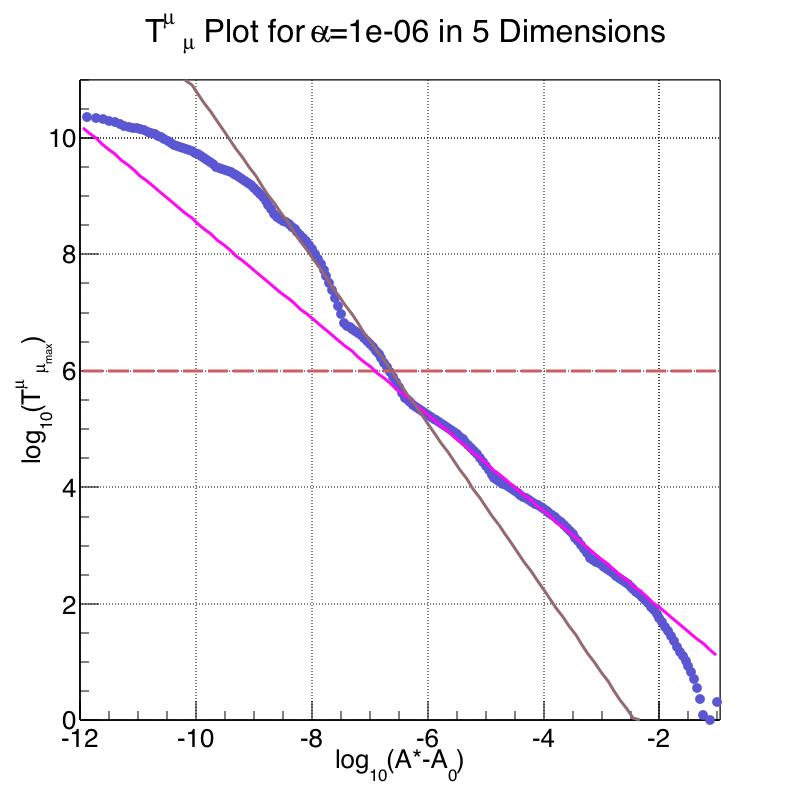}
\label{5DRicci-6}
}
\caption{$\Tm$ Scaling Plots - 5D. The lines represent the best-fit mean slopes of the curves in their respective regimes.}
\label{Ricci5D}
\end{figure}

In the case of 5D there is evidence that the slope of the radius plot decreases continuously until a minimum radius is reached, i.e. that there is a radius gap. This is most evident in Fig.(\ref{5DRicci5e-7}) but also appears to be the case in (\ref{5DRicci-6}). In the remaining 5D figures the numerics did not allow us to probe deeply enough into the GB region to fully observe this. 

A radius gap is not unexpected given the presence of the dimensionful GB parameter. Note that we focus on a radius gap instead of a mass gap because in 5D the former is trivial in light of (\ref{MAH}). 

The $\Tm$ plots, initially approximately straight, change slope quite suddenly as one moves from the GR to the GB region, and then remain constant over a small range of $\log(dA)$. The slopes are given in Table \ref{Table 1}.  As criticality is approached the slope of the $\Tm$ plot gradually decreases, suggesting that there is a maximum value to $\Tm$. This differs from GR, in which the critical solution is singular and $\Tm$ increases indefinitely as criticality is approached.

We emphasize again that these features are universal.

%%%%%%%%%%%%%%%%%%%%%%%%%%%%%%%%%%%%%%%%%%%%%%%%%%%%%%%%%%

%%%%%%%%%%%%%%%%%%%%%%%%%%%%%%%%%%%%%%%%%%%%%%%%%%%%%%%%%%%%%%
\begin{figure}[ht!]
\centering
\includegraphics[width=0.4\linewidth]{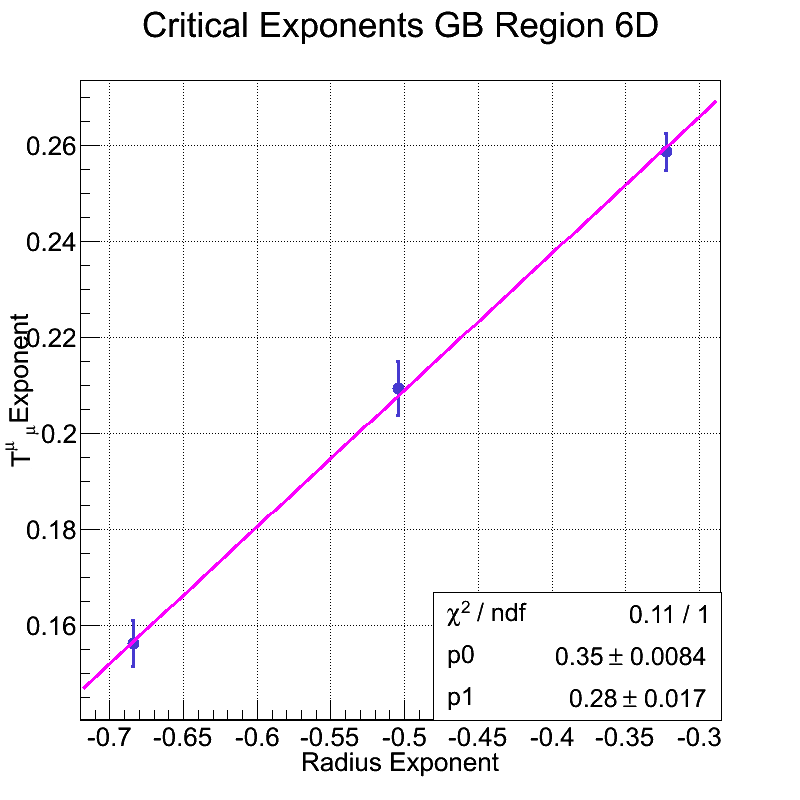}
\caption{Plot of Radius exponents vs $\Tm$ scaling exponents, 6D}
\label{Ricci vs Radius}
\end{figure}

%%%%%%%%%%%%%%%%%%%%%%%%%%%%%%%%%%%%%%%%%%%%%%%%%%%%%
%\subsubsection{6D}
In 6D things are different. There is no evidence of a radius gap in the radius scaling plots, and the slope of the $\Tm$ plots remains constant until we reach the limits of numerical accuracy. Thus it appears that there is a transition to a new set of scaling exponents, which are plotted in Table (\ref{Table 2}). Note that numerical uncertainties make the first and last entries in each column unreliable. The exponents are different for $\Tm$ and radius scaling, but the absolute value of both appear to increase with decreasing $\tilde{\alpha}$. Moreover a log-log plot of the three reliable Radius vs $\Tm$ exponents (Fig.~\ref{Ricci vs Radius}) reveals that they are related by:
\be
\gamma_{(\Tm)} \approx -(2.24\pm0.04) \times \gamma_{(Radius)}^{0.28\pm0.02}
\label{eq:Ricci vs Radius}
\ee
This is to be compared to the GR case in which the relation is determined purely by the dimension of the two quantities:
\be
\gamma_{(\Tm)} = -2 \gamma_{(Radius)}
\ee

%%%%%%%%%%%%%%%%%%%%%%%%%%%%%%%%%%%%%%%%%%%%%%%%%%%%%
\begin{figure}[ht!]
\centering
\subfigure[GR, slope$=0.43$]{
\includegraphics[width=0.4\linewidth]{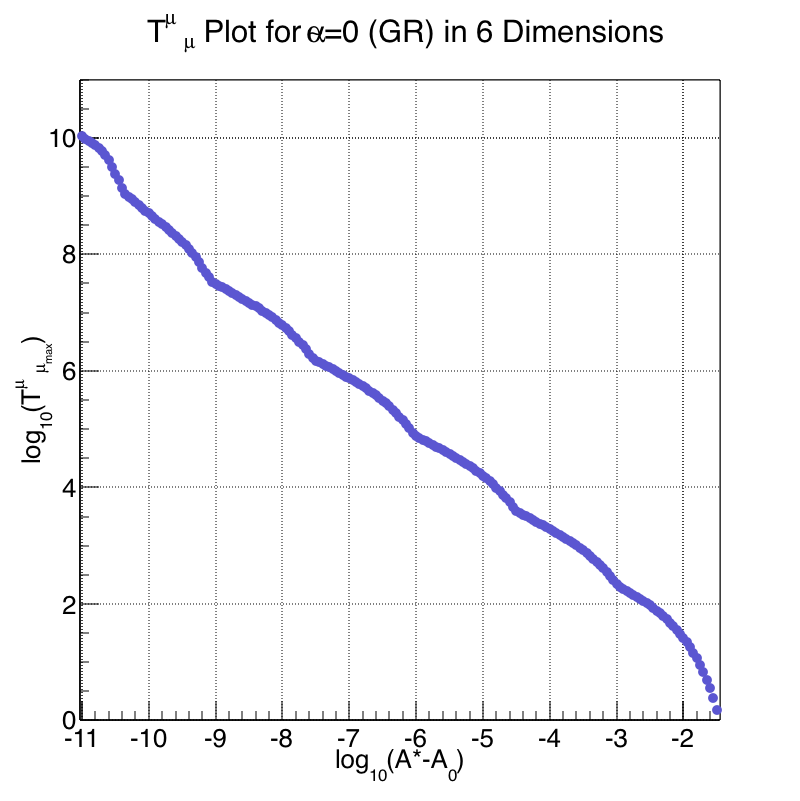}
\label{6DRicciGR}
}
\hspace{0.25in}
\subfigure[$\tilde{\alpha}=10^{-7}$]{
\includegraphics[width=0.4\linewidth]{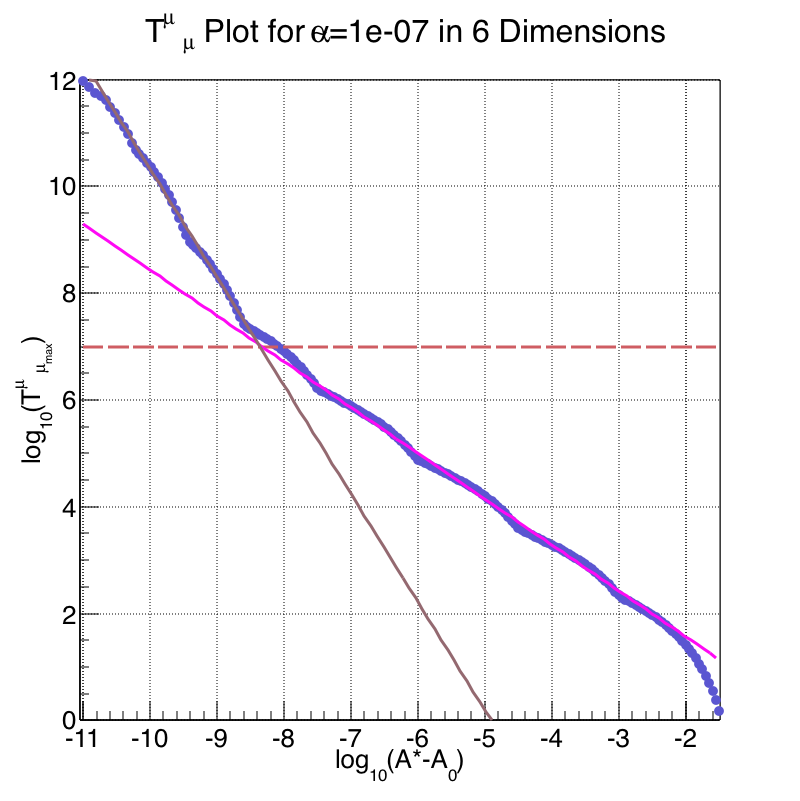}
\label{6DRicci-7}
}
\subfigure[$\tilde{\alpha}=5\times10^{-7}$]{
\includegraphics[width=0.4\linewidth]{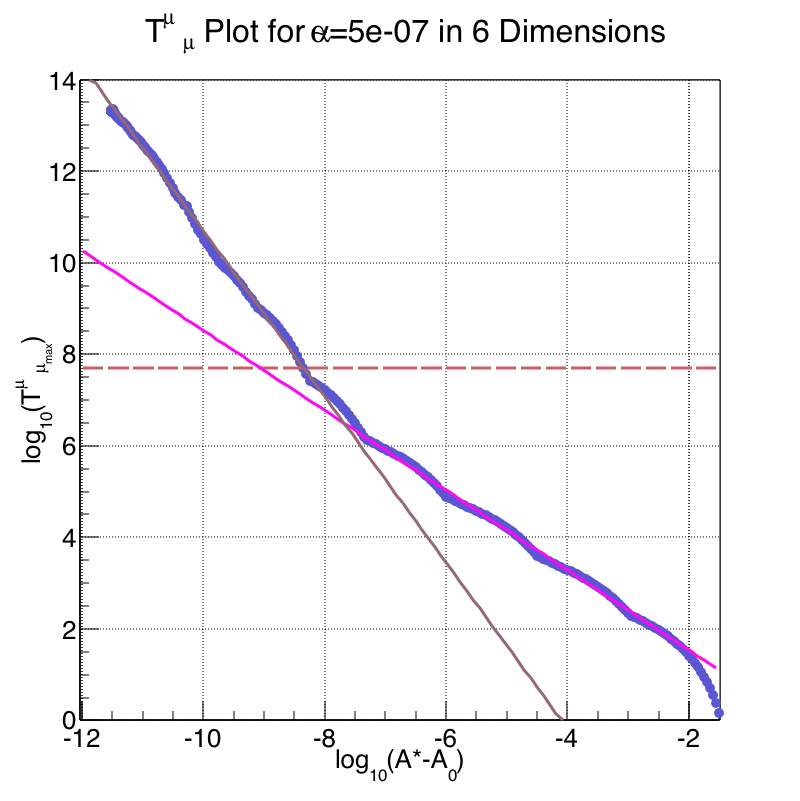}
\label{6DRicci5e-7}
}
\hspace{0.25in}
\subfigure[$\tilde{\alpha}=10^{-6}$]{
\includegraphics[width=0.4\linewidth]{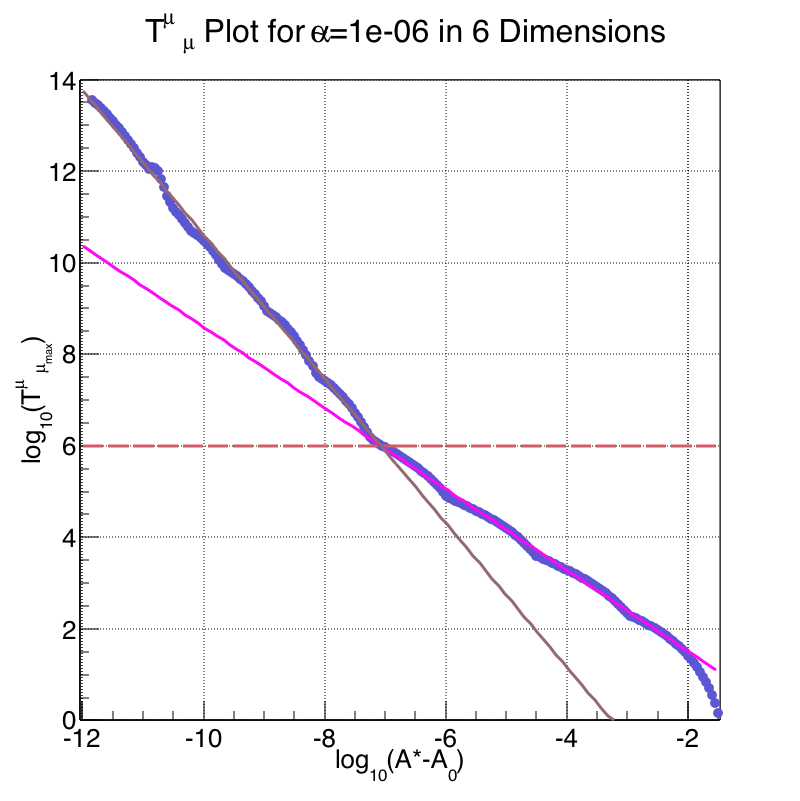}
\label{6DRicci-6}
}
\subfigure[$\tilde{\alpha}=10^{-5}$]{
\includegraphics[width=0.4\linewidth]{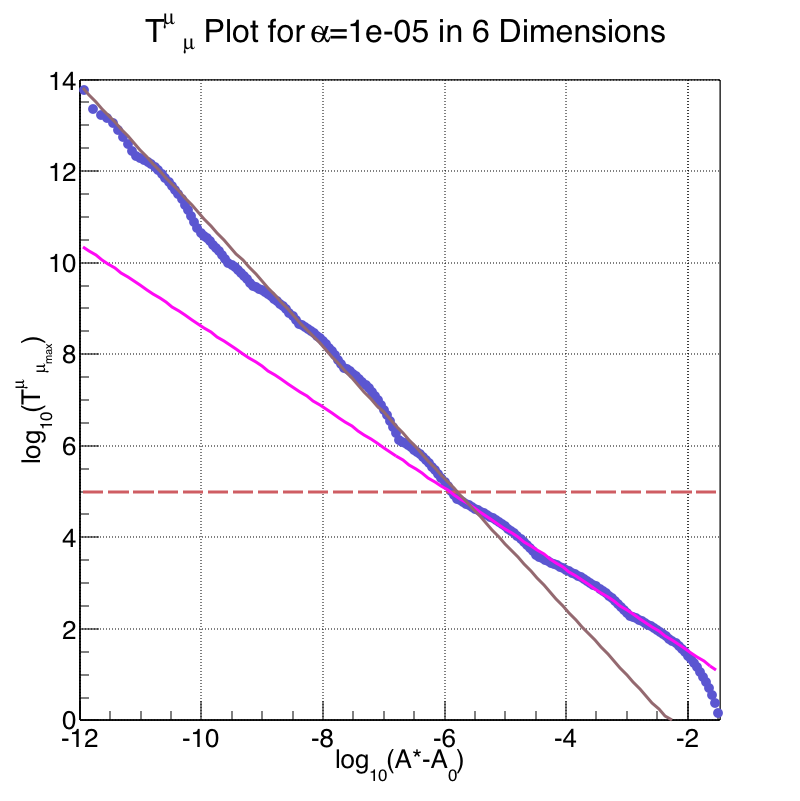}
\label{6DRicci-5}
}
\hspace{0.25in}
\subfigure[$\tilde{\alpha}=10^{-4}$]{
\includegraphics[width=0.4\linewidth]{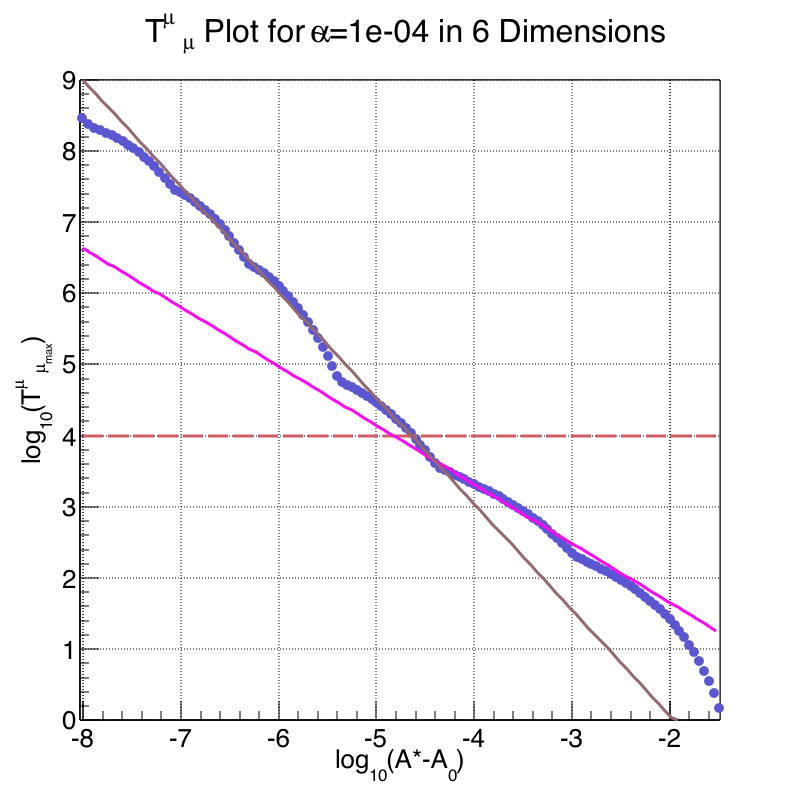}
\label{6DRicci-4}
}
\caption{$\Tm$ Scaling Plots - 6D. The lines represent the best-fit mean slopes of the curves in their respective regimes.}
\label{Ricci6D}
\end{figure}

% Tables generated by Excel2LaTeX from sheet 'GB Table'
\begin{table}
\begin{center}
\begin{tabular}{|c|c|}
\hline
%\multicolumn{ 3}{|c|}{\centering Critical Exponent in GB Region} \tabularnewline
%\hline
\centering $\tilde{\alpha}$ &  \centering 5D $\Tm$ Scaling \tabularnewline
\hline
\centering $10^{-6}$   & \centering $-1.426\pm0.074$ \tabularnewline
\hline
\centering $5\times10^{-7}$   & \centering $-1.573\pm0.076$ \tabularnewline
\hline
\centering $10^{-7}$ &  \centering $-1.577\pm0.028$  \tabularnewline
\hline
\centering $10^{-8}$  & \centering $-3.397\pm0.049$  \tabularnewline
\hline
\end{tabular}  
\caption{5D $\Tm$ scaling exponents in GB region.}
  \label{Table 1}
  \end{center}
\end{table}

%%%%%%%%%%%%%%%%%%%%%%%%%%%%%%%%%%%%%%%%%%%%%%%%%%%%%%%%%%
\begin{table}
\begin{center}
\begin{tabular}{|c|c|c|}
\hline
%\multicolumn{ 3}{|c|}{\centering Critical Exponent in GB Region} \tabularnewline
%\hline
\centering $\tilde{\alpha}$ &  \centering 6D $\Tm$ Scaling& \centering  6D Radius Scaling \tabularnewline
\hline
\centering $10^{-4}$   & \centering $-1.488\pm0.128$ &  \centering $0.257\pm0.002$ \tabularnewline
\hline
\centering $10^{-5}$   & \centering $-1.433\pm0.016$ &  \centering $0.207\pm0.002$ \tabularnewline
\hline
\centering $10^{-6}$ &  \centering $-1.619\pm0.021$ &  \centering $0.313\pm0.002$ \tabularnewline
\hline
\centering $5\times10^{-7}$  & \centering $-1.814\pm0.016$ & \centering $0.476\pm0.002$ \tabularnewline
\hline
\centering $10^{-7}$ &  \centering $-2.029\pm0.027$ & \centering \centering $0.417\pm0.002$ \tabularnewline
\hline
\end{tabular}  
\caption{6D $\Tm$ and AH radius scaling exponents in GB region.}
  \label{Table 2}
  \end{center}
\end{table}

%\caption{6D Scaling Exponents in GB region}

%%%%%%%%%%%%%%%%%%%%%%%%%%%%%%%%%%%%%%%%%%%%%%%%%%%%%%%%%
\clearpage

\section{Conclusion}
\label{Conclusion}

We studied the effects of the GB term on the dynamics of the collapse of a massless scalar field minimally coupled to gravity in five and six spacetime dimensions.  The GB term  destroys the self-similar behaviour, as demonstrated by the fact that near criticality the scalar field at the origin oscillates with a constant period. The period in five dimensions is proportional to roughly the cube root of the GB parameter and as the fourth root in six dimensions. While the 5D results differ from those in  \cite{Golod2012} it must be emphasized that we have explored a different range of GB parameter, and this may account for the difference.

We also showed the existence of modified, but still universal, horizon and $\Tm$ scaling plots near criticality. We found evidence for the existence of a radius gap in five dimensions but not in six dimensions. This qualitative difference is not completely unexpected. As mentioned below Eq.(\ref{Pisdot_code}), the time evolution equation in five dimensions is special, containing one less term than in the
higher-dimensional cases. It may also be useful to note that qualitative differences exist between five and six dimensions with regard to the stability of black holes under gravitational perturbations\cite{Gleiser2007, Soda2010}. Small five dimensional GB black holes are unstable with respect to scalar gravitational perturbations, whereas in six dimensions it is the tensor mode that yields an instability.

It is clearly of interest to confirm our results with further simulations and to try to understand analytically the source of the new scaling behaviour. 
%It should also be possible to extend the results to higher dimensions and higher order Lovelock terms. 

\section{Acknowledgements}
\label{Acknowledgements}

This work was supported in part by the Natural Sciences and Engineering Council of Canada. It also has been enabled by the use of computing resources provided by WestGrid, the Shared Hierarchical Academic Research Computing Network (SHARCNET:www.sharcnet.ca) and Compute/Calcul Canada.
 TT would like to thank the university of Manitoba for funding. We are grateful to Hideki Maeda, Dallas Clement and Patrick Brady for very informative discussions.

%\clearpage

\end{document}